%% file: edge_dipole.tex
\documentclass[10pt,prb,twocolumn,notitlepage,showpacs]{revtex4-1}
\usepackage{amsthm,amsmath,amsfonts,amssymb,verbatim, color}
\usepackage{graphicx}
\usepackage[centerlast]{caption}
\usepackage{subcaption}

\usepackage{bm}
\usepackage{multirow}

\DeclareGraphicsExtensions{.png,.jpg, .pdf}
\graphicspath{{./}}

\newcommand{\mb}{\mathbb}
\newcommand{\mf}{\mathfrak}
\newcommand{\mc}{\mathcal}

\begin{document}
\title{Guiding-center Hall viscosity and intrinsic dipole moment \\ along edges of incompressible fractional quantum Hall fluids}
\author{YeJe Park and F. D. M. Haldane}
\affiliation{Department of Physics, Princeton University,
Princeton NJ 08544-0708}

\date\today

\begin{abstract}
The discontinuity of guiding-center Hall viscosity (a bulk property) at edges of incompressible quantum Hall fluids
is associated with  the presence of an intrinsic electric dipole moment on the edge. If there is a gradient of
drift velocity due to a non-uniform electric field, the discontinuity in the induced stress is exactly balanced by
the electric force on the dipole.
The total Hall viscosity has two distinct contributions: a ``trivial'' contribution associated with the
geometry of the Landau orbits, and  a non-trivial contribution associated with guiding-center correlations.
   We  describe a relation between
the guiding-center edge-dipole moment and  ``momentum polarization'',  which relates the guiding-center part of the
bulk Hall viscosity to the ``orbital entanglement spectrum(OES)''.
We  observe that using the computationally-more-onerous
``real-space entanglement spectrum (RES)''  just adds the trivial
Landau-orbit contribution to the guiding-center part. This shows that
all the non-trivial
information is completely contained in the  OES, which also exposes a
fundamental topological quantity $\gamma$ = $\tilde c-\nu$, the difference between the ``chiral stress-energy anomaly'' (or signed conformal anomaly) and the chiral
charge anomaly.  This quantity characterizes correlated  fractional
quantum Hall fluids, and vanishes in uncorrelated  integer quantum Hall fluids.
\end{abstract}
\maketitle

\input{./intro}

\input{./gc_hall_vs}

\section{Numerical method.1 : \\ Jack polynomials }
\label{sec:methods}

\label{sec:jack}

In the last section, we obtained an expression (\ref{gc_spin_and_dipole}) relating the guiding-center spin and the intrinsic dipole moment. To find the guiding-center spin, we assumed that the shape of the QH fluid was a droplet. It is not yet clear if the expression (\ref{gc_spin_and_dipole}) is independent of the shape of the QH fluid. In this section, we will find the exact model FQH states(Laughlin and Moore-Read states) on a cylinder from Jack polynomials. We will then calculate their intrinsic dipole moments, and confirm the correctness of (\ref{gc_spin_and_dipole}) for a straight edge.

In Sec.\ref{sec:mapping}, we first describe how one can obtain the many-particle incompressible model FQH state wavefunction from  symmetric polynomials known as Jack polynomials\cite{Bernevig2008}.

Then, in \ref{sec:Fermi}, we discuss how we identify the analogs of Fermi momenta for finite-size FQH ground states on a cylinder. The precise identification of the Fermi momenta is necessary to obtain the correct value of the intrinsic dipole moment.

In \ref{sec:jack_occ},  we calculate and plot the occupation number profiles for the model FQH wavefunctions obtained from Jacks, and we compared them with the behavior predicted by the chiral boson theory\cite{Wen1990}.

In \ref{sec:luttinger}, we show the validity of the Luttinger's theorem in application to incompressible FQH states.

 In \ref{sec:dipole}, we calculate the intrinsic dipole moment, and confirm (\ref{gc_spin_and_dipole}) which was first predicted by Haldane\cite{Haldane2009}.

\input{./jack}

\section{Numerical Method.2 : Entanglement Spectrum}
\label{sec:es}

In Sec.\ref{sec:es_dipole}, we introduce the orbital entanglement spectrum(OES)\cite{Li2008}.
We observe that the chirality of the OES can be explained by the fact the model states derives from Jack polynomials.
Then, we describe how to calculate  the total net momentum quantum number(``momentum polarization") for a subsystem on a cylinder.
We also  derive the minimum change in the momentum quantum number as a function of the change in the particle number within a subsystem from the manipulation of root occupation numbers.

 In Sec.\ref{sec:decomposition}, we relate the momentum polarization with the intrinsic dipole moment. We show that the momentum polarization can be decomposed into three distinct parts.  We also show that one of the three which known as ``topological spin" can be calculated solely from the root occupation numbers. The other topological term $\gamma = \tilde c - \nu$ is identified as a purely FQHE quantity which vanishes for IQHE.

 In \ref{sec:ent_realsp}, we show that the momentum polarization calculated from the RES merely adds  a trivial Landau-orbit contribution to the one calculated from the OES.

\input{./oes}

\input{./res}

\input{./conclusion}

\input{./edge_dipole.bbl}
\end{document}

%% file: intro.tex
\section{Introduction}

During the period of three decades since the first observation\cite{Tsui1982}, incompressible fractional quantum Hall (FQH) states have been shown to possess  many intriguing properties, and one of these is the ``Hall viscosity"\cite{Avron1995,Read2009,Read2011,Haldane2009}. The total Hall viscosity tensor $\eta'_H \,^{ab}_{cd}$ of an incompressible FQH state is a sum of two parts of different origins,
\begin{equation*}
\eta_H'\,^{ab}_{cd} =  \tilde {\eta}_H\,^{ab}_{cd} + \eta_H\,^{ab}_{cd}.
\end{equation*}
The former, ``Landau-orbit Hall viscosity" $\tilde \eta_H \,^{ab}_{cd}$, is the response to the variation of the shape of the Landau-orbit(i.e. cyclotron motion), and the latter, ``guiding-center Hall viscosity" $\eta_H \,^{ab}_{cd}$, is the response to the variation of the shape of the correlation hole.  Each of these two shapes can be parameterized by a 2$\times$2 spatial  metric\cite{Haldane2011}. The metric associated with the Landau-orbit shape is called ``Landau-orbit metric" and the one associated with the correlation hole shape is called ``guiding-center metric". The generalization by Haldane\cite{Haldane2011}  to dynamical  variation of the guiding-center metric led to a new research interest for the geometric description of incompressible FQH states\cite{Son2012,Maciejko2013}.

 There have been attempts to link the Hall viscosity with other physical observable such as the Hall conductivity \cite{Son2012, Bradlyn2012}. The basic assumption of those calculations is Galilean invariance. However, an incompressible FQH state is a topological phase for which such assumption should  not be essential.
 In this report, we relate the guiding-center Hall viscosity, i.e. the part  of the Hall viscosity due to the guiding-center degrees  of freedom, with the intrinsic dipole moment per unit length along the edge of incompressible FQH states\cite{Haldane2009}. Though we will do the computation for a straight edge, the result is applicable to the edge of an arbitrary shape because the relationship derives from the local force balance at a  point on the edge.

Another aim of this report is to show that we can calculate the intrinsic dipole moment from the ``orbital entanglement spectrum" (OES)\cite{Li2008}. Therefore, OES contains enough information to determine the guiding-center Hall viscosity.
The intrinsic dipole moment  is essentially the non-vanishing mean momentum due to the entanglement with the other half of the whole system (also called ``momentum polarization").
For a finite length $L$ of the edge, there is a correction of order $\mc O(L^0)$.  This correction is composed of two parts, ``topological spin"\cite{Zaletel2012, Qi2012} and a new topological quantity $\gamma = \tilde c - \nu$ which is the difference between \textit{signed} conformal anomaly $\tilde c$ and chiral charge anomaly $\nu$ of the underlying edge theory\cite{Wen1990}. We also elucidate the origins of the topological spin and the fractional charge by showing that they originate from different cuts relative to the ``root occupation pattern"\cite{Bernevig2008}.

The last goal of this  report is to show that the computation of Hall viscosity with the so-called ``real-space entanglement spectrum" (RES)\cite{Dubail2012,Zaletel2012} merely adds  ``Landau-orbit Hall viscosity" which is a rather trivial part of the Hall viscosity due to the cyclotron motion.  For a finite length $L$, RES also adds  the chiral anomaly $\nu$ to the $\mc O(L^0)$ correction, and therefore obscures the existence of the new topological quantity $\gamma$. We are led  to  claim that all the essential information of FQH states are contained in OES.
\\

\begin{figure}
        \centering
        \begin{subfigure}[b]{0.23\textwidth}
                \includegraphics[width=\textwidth,page=2]{./figures}
                \caption{A straight edge}
                \label{velocity}
        \end{subfigure}
       ~
        \begin{subfigure}[b]{0.2\textwidth}
                \includegraphics[width=\textwidth]{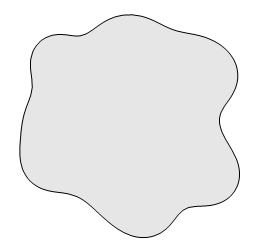}
                \caption{An arbitrary edge}
                \label{arbitrary_edge}
        \end{subfigure}
        \caption{The gray area represents the Hall fluid. In general, the drift velocity depends on the distance from the edge. Each edge follows an equipotential line.}\label{fig:animals}
\end{figure}

First of all, let's clarify what we mean by the intrinsic dipole moment.  Consider electrons on a cylinder through whose surface an uniform magnetic field  $B$ passes. We confine the electrons  by an external electric potential $V(y)$ that depends on $y$, one of the two spatial coordinates, $x$ and $y$. Then, single-particle states $|\phi_m\rangle$ are labeled by guiding-centers $y_m = 2\pi m \ell_B^2 / L$, $m  \in \mb Z +\textstyle{\frac{1}{2}}$ ($\ell_B^2 = \hbar/ eB$).
Given a many-particle state $|\Psi\rangle$, we can calculate its occupation-number profile which is the set of the expectation values of occupation-number operators $n_m$ for each index $m$.
For instance, consider   an IQH state in the first Landau level, $|\Psi_1\rangle$, filling the upper-half plane with a ``Fermi momentum" at $y = 0$ (See Fig.\ref{velocity}). Then, its occupation profile is $\{\dots, n_{-3/2},n_{-1/2},n_{1/2},n_{3/2},\dots\} = \{\dots,0,0,\nu,\nu,\dots\}$ where the filling factor $\nu = 1$. In the continuum limit $L\rightarrow \infty$, the occupation profile for this uncorrelated state is a step function in $y$ : $n(y) = \nu \,\theta(y)$.

\begin{figure}
\includegraphics[width=0.8\linewidth,page=1]{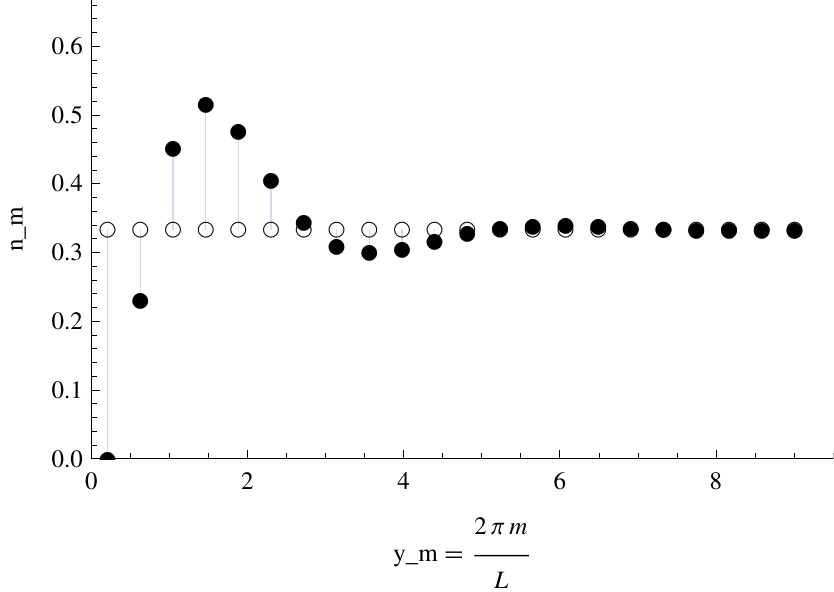}
\caption{Occupation profile ($\bullet$) for Laughlin $\nu = 1/3$ state and the uniform occupation profile $\bar n_m = \nu$ ($\circ$)  on a cylinder with circumference $L = 15$ ($\ell_B = 1$).}
\label{occupation_example}
\end{figure}

 Now, let's consider as an example of a correlated state, the Laughlin $\nu = \textstyle{\frac{1}{3}}$ state\cite{Laughlin1983}, $|\Psi_{1/3}\rangle$. As before, suppose the Fermi momentum is at $y = 0$ (when the circumference $L$ is finite, it is not obvious where the Fermi momentum is. This will be clarified later, Sec.\ref{sec:Fermi}). For a given $L$, we can obtain a occupation profile. For $L = 15 \ell_B$, we have the occupation profile in Fig.\ref{occupation_example}. Unlike the uncorrelated state $|\Psi_1\rangle$, the occupation profile of the correlated state $|\Psi_{1/3}\rangle$ deviates from the filling factor $\nu = \textstyle{\frac{1}{3}}$ near the edge. In the continuum limit, the occupation profile becomes $n(y) \propto y^{(\nu^{-1}-1)}$ as predicted by chiral boson theory\cite{Wen1990} (cf. Sec.\ref{sec:jack_occ}).
 We see that  the correlation among the electrons develops an extra ``intrinsic dipole moment" at the edge by ``pulling them inward" (this corresponds to the fact that FQH model wavefunctions are spanned by states obtainable by ``squeezing" the ``root state"\cite{Bernevig2008}, See Sec.\ref{sec:mapping}).

Because an incompressible FQH state is a topological phase,  the straightness of the edge should not be essential. Therefore, we consider an edge of an arbitrary shape as in Fig.\ref{arbitrary_edge} on a flat 2D plane. We denote the line element along the edge by $d L^a$.  We relate the intrinsic dipole moment $d p^a$ per a line element $d L^a$ by introducing a dimensionless symmetric 2-tensor $Q^{ab}$,
 \begin{equation*}
 dp^a = -e\,Q^{ab} \epsilon_{bc} dL^c.
 \end{equation*}
The electric charge $-e$ is negative, and $\epsilon_{ab}= \epsilon^{ab}$ is the Levi-Civita anti-symmetric tensor, $\epsilon_{xy} = - \epsilon_{yx} = 1$. Throughout this report, we distinguish covariance and contravariance of indices, and we use the Einstein summation convention.  In general, the electric field which derives from the Coulomb interaction and the confining potential is not constant but depends on the distance from the edge. The gradient of the electric field coupled with the intrinsic dipole moment results in an electric force,
 \begin{equation*}
 d F_{\text{el},a} =  dp^b \partial_a E_b.
 \end{equation*}
If the edge is to be stable, this electric force should be balanced. What should this counter-balancing force be? The counter-balancing force against the electric force on the intrinsic dipole comes from the guiding-center Hall viscosity. Here, we review the physical argument\cite{Haldane2009}, and then we will provide two kinds of numerical proofs, first utilizing  the exact model wavefunctions  in Sec.\ref{sec:jack} and secondly utilizing the orbital entanglement spectra in Sec.\ref{sec:es_dipole}.

Firstly, we note that  pressure is absent. An incompressible FQH state is a topological quantum phase. In the bulk, all excitations are separated by an energy gap, and its low-energy effective description is the Chern-Simons  Lagrangian\cite{Zhang1989} with a vanishing Hamiltonian. Because the incompressible state has no phonons to mediate the effect of external force, the bulk pressure vanishes entirely\cite{Haldane2009}.

We should take into account only the guiding-center part of the total Hall viscosity because the ``trivial" Landau-orbit Hall viscosity $ \tilde {\eta}_H\,^{ab}_{cd} $ is present whether or not the electrons are correlated (this part of the Hall viscosity will be discussed together with RES in Sec.\ref{sec:ent_realsp}).
When the electrons develop correlations among themselves, there arises the additional non-trivial guiding-center Hall viscosity $ \eta_H\,^{ab}_{cd}$, concurrently with the intrinsic dipole moment.

The non-uniform electric field near the edge results in a non-vanishing gradient of the drift velocity $v^a = \epsilon^{ab} E_b/B$.
Then, the edge experiences a dissipationless stress $\sigma_a^b$ due to the guiding-center Hall viscosity proportional to the gradient of the drift velocity,
\begin{equation*}
\sigma^a_b = -\eta_H \,^{ac}_{bd}\, \partial_c v^d
 = -\eta_H^{aecf} \epsilon_{eb}\epsilon_{fd} \,\partial_c v^d.
\end{equation*}
where in the second equality, we raised the two lower indices of $\eta_H\,^{ac}_{bd}$ using Levi-Civita tensors. Note that $\eta_H^{abcd}$ is anti-symmetric under the exchange of the two pairs of indices $(ab) \leftrightarrow (cd)$, and symmetric under the exchange of two indices $ a\leftrightarrow b$ or $c \leftrightarrow d$ (cf. Sec.\ref{sec:hall_vs}). Such 4-tensor can be expanded in terms of a symmetric 2-tensor $\eta_H^{ab}$,
\begin{align*}
\eta_H^{abcd}&= \textstyle{\frac{1}{2}}(\epsilon^{ac}\eta_H^{bd} + \epsilon^{ad}\eta_H^{bc}+ a\leftrightarrow b).
\end{align*}
With this expansion, the expression for the stress tensor becomes
\begin{equation*}
\sigma_b^a = \textstyle{\frac{1}{2}}B^{-1}(\epsilon^{ac} \epsilon_{eb} \eta_H^{ef} + \delta^c_b \eta_H^{af} + c \leftrightarrow f ) \,\partial_c E_f
\end{equation*}
From the stress, we find the dissipationless viscous force $d F_{\text{visc},a}$ on a line element $d L^a$,\cite{Landau1986}
\begin{equation*}
d F_{\text{visc},a} = \sigma_a^b \epsilon_{bc} d L^c.
\end{equation*}
We make two physical assumptions to reduce the viscous force equation further. The first assumption is that the magnetic field is static so that the Maxwell's equation gives $\epsilon^{ab}\partial_a E_b = 0$. This implies the 2-tensor $\partial_a E_b$ is symmetric under the exchange of the indices, $a \leftrightarrow b$. The second assumption is that the line element $d L^a$ of the edge is directed along the equipotential line so that $ E_a d L^a = 0$. From these two assumptions, the viscous force reduces to
\begin{equation*}
d F_{\text{visc},a} =   (B^{-1} \eta^{bc}_H \epsilon_{cd} d L^d )\partial_a E_b .
\end{equation*}
Then, from the requirement that the net force on the line element  vanishes, $d F_{\text{el},a} + d F_{\text{visc},a}= 0$, we obtain the relationship between the intrinsic dipole moment tensor $Q^{ab}$ and the guiding-center Hall viscosity 2-tensor $\eta_H^{ab}$,
\begin{equation*}
 \eta_H^{ab} = e B Q^{ab} .
\end{equation*}
Thus, if we know the guiding-center Hall viscosity tensor $\eta_H^{ab}$, we also know the intrinsic dipole moment $d p^a$  along the static equipotential edge,
\begin{equation}
d p^a = -  B^{-1} \eta^{ab}_H \epsilon_{bc} d L^c.\label{main}
\end{equation}
This relationship (\ref{main}) between the intrinsic dipole moment and the guiding-center Hall viscosity tensor was derived from a local balance of forces. Therefore, we expect the relationship to hold for an edge of any smooth arbitrary shape reflecting the topological nature of an incompressible FQH state.

To give a specific example, consider the situation depicted  in Fig.\ref{velocity} for which $\partial_y v^x$ is the only non-vanishing component of the velocity gradient. Then, the stress expression reduces to
\begin{equation*}
\sigma^y_y = -  \eta^{yy}_H \frac{\partial_y E_y}{B}.
\end{equation*}
The viscous force per a line element $d L^x$ on the edge is given by
\begin{equation*}
d F_{\text{visc},y} = \eta^{yy}_H \frac{\partial_y E_y}{B} dL^x.
\end{equation*}
The electric force on the dipole for this situation is
\begin{equation*}
d F_{\text{el},y} = -e\,Q^{yy} dL^x \partial_y E_y.
\end{equation*}
Vanishing of the net force gives us,
\begin{equation}
\frac{ dp^y}{dL^x} = \frac{\eta^{yy}_H}{B}  . \label{main_straight}
\end{equation}

The left-hand side of (\ref{main_straight}) can be numerically calculated from occupation profiles (for instance, Fig.\ref{occupation_example}). This will be done in Sec.\ref{sec:jack} and Sec.\ref{sec:es}. The right-hand side of (\ref{main_straight}) can be analytically calculated as the expectation value of the ``area-preserving deformation generators". This calculation is described in Sec.\ref{sec:def}.

%% file: gc_hall_vs.tex
\section{Theoretical Background}

\label{sec:def}

In this section, we describe the distinct physical origins of  the Landau orbit metric and the guiding-center metric. The Landau-orbit Hall viscosity and the guiding-center Hall viscosity  are derived as the adiabatic responses to the variation of the Landau orbit metric and the guiding-center metric respectively, and as the expectation values of area-preserving deformation generators.
We calculate these quantities for the model Laughlin\cite{Laughlin1983} and Moore-Read\cite{Read1991} states.

\subsection{Landau-orbit metric and guiding-center metric}

Consider $N$ electrons with charge $- e <0$ living on a 2D plane subject to a normal magnetic field strength $\bm B = B \hat z$, $B>0$. The $i$-th electron  on the 2D plane  has four degrees of freedom, its coordinate $\bm r_i$ and its dynamical momentum $\bm \pi_{i} =\bm p_{i} + e \bm A(\bm r_i)$. Note that $i,j,\dots$ are indices for electrons, and $a,b,\dots$ are indices for the spatial coordinates. The coordinate operator can be decomposed into two  operators
\begin{equation}
\bm r_i= \bm R_i + \tilde{\bm  R_i},
\end{equation}
where the first operator $\bm R_i$ is the ``guiding-center of the electron, and the second operator $\tilde{\bm  R_i}$ is  the ``Landau-orbit radii".
The Landau-orbit radii is defined in terms of the dynamical momenta: $\tilde R^a_i = \epsilon^{ab}\pi_b/ e B$.
 These operators have the following commutation relations,
\begin{subequations}
\begin{align}
[\tilde R_i^a , \tilde R_j^b]& =- i \epsilon^{ab} \delta_{ij} \ell_B^2\\
[ R_i^a ,  R_j^b]& = i \epsilon^{ab} \delta_{ij} \ell_B^2\\
[\tilde R_i^a , R_j^b]& = 0,
\end{align}
\end{subequations}
where $\ell_B^2 = \hbar / eB$. This decoupling between $\bm R_i$ and $\tilde{\bm R}_i$ is completely independent of the choice of a gauge.

Out of these operators, we can form area-preserving deformation (APD) generators\cite{Haldane2011}
\begin{subequations}
\begin{align}
\Lambda^{ab} & = \textstyle{\sum_i} \Lambda^{ab}_i = \textstyle{ \frac{1}{4\ell_B^2}}\sum_i \{ R_i^a, R_i^b\}\\
\tilde \Lambda^{ab} & = \textstyle{\sum_i} \tilde \Lambda^{ab}_i = \textstyle{ \frac{1}{4\ell_B^2}}\sum_i \{ \tilde R_i^a, \tilde R_i^b\},
\end{align}
\end{subequations}
where $\{ \, , \, \}$ is an anti-commutation. These satisfy the commutation relations of $\mf{sl}(2,\mb R) $
\begin{subequations}
\begin{align}
[\Lambda^{ab}, \Lambda^{cd}] & = +\textstyle{\frac{i}{2}} (\epsilon^{ac}\Lambda^{bd} + \epsilon^{ad} \Lambda^{bc} + a \leftrightarrow b)\\
[\tilde \Lambda^{ab}, \tilde \Lambda^{cd}] & = - \textstyle{\frac{i}{2}} (\epsilon^{ac}\tilde \Lambda^{bd} + \epsilon^{ad}\tilde \Lambda^{bc} + a \leftrightarrow b).
\end{align}\label{APD_comm}
\end{subequations}

The interacting electrons are described by the following Hamiltonian $H$ which is a sum of a single-particle energy $H_0$ and the interaction $V$,
\begin{align}
H & = H_0 + V  \nonumber \\
V & =  \textstyle{ \frac{1}{2}}\sum_{i \neq j} V(  \bm r_i - \bm  r_j; \varepsilon) \label{full_hamiltonian}.
\end{align}
Note that the Coulomb interaction $V$ also depends on the permittivity tensor $\varepsilon_{ab}$. The most general single-particle energy is
\begin{equation*}
H_0 = \textstyle{\sum_i} h(\tilde{\bm R}_i),
\end{equation*}
where $h(\bm r)$ is a function of $\bm r$ whose constant contours are non-overlapping and closed.
The most general form of the single-particle energy  is technically intractable, so we take a model single-particle energy parameterized by a unimodular symmetric positive-definite 2-tensors $\tilde g_{ab}$ which we call the ``Landau-orbit metric",
\begin{equation}
H_0 = \textstyle{\sum_i} h(\tilde g_{ab} \tilde R_i^a \tilde R_i^b),\label{simplified_single_particle_energy}
\end{equation}
where $h(r)$ is a monotonically increasing function of $r$. This form includes, for instance, the following two examples which break Galilean invariance,
\begin{align*}
H_0 & = \textstyle{\sum_i}(\tilde g_{ab} \Lambda_i^{ab})^{k+1}, \quad k \in \mb N\\
H_0 & = \textstyle{\sum_i}\sqrt{ 1+ \tilde g_{ab} \Lambda_i^{ab}},
\end{align*}
The second example  is the massive Dirac Hamiltonian of a charged particle subject to a normal magnetic field.

 If the system is Galilean invariant, the  Landau-orbit metric $\tilde g_{ab}$ is determined by the effective mass tensor $(m^{-1})^{ab}$,
\begin{align*}
H_0 & = \textstyle{ \frac{1}{2}}( m^{-1})^{ab}\sum_i \pi_{i,a} \pi_{i,b}  \nonumber \\
&= \textstyle{\frac{1}{2}}\hbar \omega_c\, \tilde L( \tilde g)\\
\tilde L(\tilde g)
& =\textstyle{\sum_i} \tilde L_i(\tilde g)
=\textstyle{ \sum_i}\tilde g_{ab}\tilde \Lambda_i^{ab} ,
\end{align*}
where the cyclotron frequency is $\omega_c = e B /|m|$ and $| m| = \det m$. We also defined the rotation generator of Landau-orbit radii, $\tilde L(\tilde g)$.  The eigenvalues of $\tilde L_i (\tilde g)$ are $\tilde s_n = n+\textstyle{\frac{1}{2}}, \, n \in \mb Z_+$, and we call $\tilde s_n$ the ``Landau-orbit spin".

For the single-particle energy (\ref{simplified_single_particle_energy}), we can label the Landau level with the non-negative integer  $n$ from the eigenvalue of $\tilde L_i(\tilde  g)$. Note that the system without Galilean invariance has an unequal energy gap between neighboring Landau levels.

In the strong magnetic field strength limit where Landau level mixing is not allowed, the Landau-orbit and guiding-center degrees of freedom decouple. Then, the  many-particle ground state $|\Psi\rangle$ of the Hamiltonian $H$ in the $n$-th Landau level can be decomposed as a tensor product,
\begin{equation}
|\Psi \rangle =   (\textstyle{\prod_i} |\psi_{i,n}\rangle_L) \otimes |\Psi (g)\rangle_G .\label{wf_decomp}
\end{equation}
The vectors with the subscript $L$ (for Landau-orbit) can be acted on only by the Landau-orbit operators $\tilde {\bm R}_i$ and the vectors with the subscript $G$ (for guiding-center) can be acted on only by the guiding-center operators $\bm R_i$. The vector $|\psi_{i,n}\rangle_L$ is the $n$-th eigenstate of $\tilde L_i(\tilde g)$.

We now discuss how the guiding-center part $|\Psi(g)\rangle_G$ is determined. Since the Landau-orbit part of $|\Psi\rangle$ is fixed, the interaction $V$ may be projected into the $n$-th Landau level,
\begin{equation}
 \Pi_n V \Pi_n = \frac{1}{2 N_{\Phi}} \sum_{\bm q} V(\bm q ;\varepsilon)  f_n(\bm q )^2 \rho(\bm q) \rho(- \bm q)\label{projected_hamiltonian},
\end{equation}
where $\Pi_n$ is the projection operator into the $n$-th Landau level, $N_{\Phi}$ is the total number of flux quanta penetrating the QH fluid, $V(\bm q ; \varepsilon)$ is the Fourier-transformation of $V(\bm r ; \varepsilon)$. We define the ``Landau-orbit form factor" $f_n(\bm q )$ and the guiding-center density operator $\rho(\bm q)$ as follows. Consider the Fourier-transformation of the density operator $\rho_0(\bm r)$,
\begin{equation*}
\rho_0(\bm q) = \textstyle{\sum_i} e^{i \bm q \cdot \bm r_i}.
\end{equation*}
The density operator $\rho_0(\bm q)$ is projected into the $n$-th Landau level by sandwiching the operator with the vector $\textstyle{\prod_i} |\psi_{i,n}\rangle_L$, and this produces the projected density operator $ \rho_n(\bm q )$ as a product of the form factor and the guiding-center density operator,
\begin{subequations}
\begin{align}
 \rho_n(\bm q ) & = f_n(\bm q) \rho(\bm q)\\
f_n(\bm q ) & = \langle \psi_{i,n}| e^{i\bm q\cdot \tilde {\bm R}_i} |\psi_{i,n}\rangle_L \\
\rho(\bm q) & = \textstyle{\sum_i} e^{ i \bm q\cdot \bm R_i}.
\end{align}
\end{subequations}
If the single-particle energy is of the form (\ref{simplified_single_particle_energy}), the Landau-orbit form factor becomes
\begin{equation}
 f_n(\bm q) = L_n\left( \textstyle{\frac{1}{2}}|\bm q|_{\tilde g}^2\right)\, e^{-|\bm q|_{\tilde g}^2/4} \label{landau_form}
 \end{equation}
where $L_n $ is a Laguerre polynomial of degree $n$, and the Landau-orbit metric norm is defined as $|\bm q|_{\tilde g}^2 = \tilde g^{ab} q_a q_b\ell_B^2$ ($\tilde g^{ac} \tilde g_{cb} = \delta^a_b$). We can make an alternative definition of the Landau-orbit metric in terms of the Landau-orbit form factor,
\begin{equation}
\tilde g^{ab} = \tilde s_n^{-1}\ell_B^{-2} \partial_{q_a}\partial_{q_b} f_n(\bm q)|_{\bm q = \bm 0}.
\end{equation}
This definition gives us the interpretation of the Landau-orbit metric as the parameter which determines the shape of the Landau-orbit.

One can re-write the projected interaction (\ref{projected_hamiltonian}) into a more fundamental expansion known as Haldane pseudo-potential \cite{Haldane1983},
\begin{subequations}
\begin{align}
\Pi_n V \Pi_n & = \sum_{m= 0}^{\infty} V_m( g, n ,\varepsilon) P_m( g)  \label{pseudo}\\
V_m(g,n,\varepsilon) &= \frac{1}{2N_{\phi}} \sum_{\bm q} V(\bm q;\varepsilon) f_n(\bm q)^2  L_m(|\bm q|_g^2)e^{-|\bm q|_g^2}\\
P_m(g)& =\frac{1}{N_{\phi}}\sum_{\bm q} L_m(|\bm q|_g^2)  e^{-|\bm q|_g^2/2} \rho(\bm q) \rho(-\bm q).
\end{align}
\end{subequations}
Here, we introduced a positive-definite symmetric 2-tensor $g_{ab}$ which we call the ``guiding-center metric"  through the norm $|\bm q|_g = g^{ab} q_a q_b \ell_B^2$ ($g^{ac} g_{cb} = \delta^a_b$). $P_m(g)$ is a projection operator: to understand the action of $P_m(g)$, we  define the ``relative guiding-center rotation generator",
\begin{equation}
L_{ij}(g) = \textstyle{\frac{1}{8\ell_B^2}} g_{ab} \{R^a_i - R^a_j, R^b_i -R^b_j\}.
\end{equation}
This operator has a spectrum, ``relative guiding-center angular momentum" $\{ m +\textstyle{\frac{1}{2}} : m \in \mb Z_+\}$. $P_m(g)$ has non-vanishing matrix elements for states with a pair of particles with relative guiding-center angular momentum $m +\textstyle{\frac{1}{2}}$.

 Instead of using the full expansion as given in (\ref{pseudo}), we can form a model interaction,
\begin{equation}
V_{\text{model}}(g) = \sum_{ m = 0}^{q-1} V_m P_m(g), \label{model_int}
\end{equation}
where $V_m$ are positive reals and $q$ is some positive integer. The full expansion (\ref{pseudo}) does not depend on a particular choice of $g_{ab}$. However, the model interaction does depend on $g_{ab}$. The $\nu = \textstyle{\frac{1}{q}}$ ``Laughlin state" is an exact zero energy state of the model interaction ($V_{\text{model}}(g)|\Psi(g)\rangle_G = 0$). The Laughlin wavefunction is a particular member ($g_{ab} = \tilde g_{ab}$) of the family of Laughlin states parameterized by $g_{ab}$ in the Galilean invariant system\cite{Haldane2011}. If the original projected interaction (\ref{projected_hamiltonian}) contains the permittivity tensor $\varepsilon_{ab}$ and the Landau-orbit metric $\tilde g_{ab}$ that are not related by multiplying a constant, then there is no reason to prefer the isotropic state with $g_{ab} = \tilde g_{ab}$.

Therefore, we see that the particular form of the model interaction (i.e. the set of numbers $\{V_m\}$) determines the correlation among particles; it tells us what relative guiding-center momenta are energetically unfavorable. Meanwhile, the guiding-center metric determines the shape of the correlation hole.

Given the family of states  $\{|\Psi(g)\rangle_G : g_{ab} \}$ which minimize $V_{\text{model}}(g)$ and are parameterized by $g_{ab}$, the equilibrium guiding-center metric is finally determined by minimizing the correlation energy,
\begin{equation}
E_G(g) = \langle \Psi(g)| \Pi_n V \Pi_n |\Psi (g)\rangle_G. \label{correlation_energy}
\end{equation}
Note that the guiding-center metric describes an emergent geometry of the correlated electrons while Landau-orbit metric directly comes from the Landau-orbit form factor. The guiding-center metric may vary on the length scale much larger than $\ell_B$. Furthermore, it was proposed by Haldane\cite{Haldane2009} to be a dynamical field that describes the gapped collective mode of the incompressible FQH fluid.

\subsection{Hall viscosity}

\label{sec:hall_vs}
In the last section, we described the definition of the equilibrium values of the Landau-orbit metric and the guiding-center metric. Here, we want to deform the metrics preserving their determinants, and find the Hall viscosities as the response of the incompressible FQH state without assuming Galilean and rotation invariances.

The APD generators preserve the determinant of the metric $g_{ab}$ and $\tilde g_{ab}$. To see this (let's focus on guiding-centers first), define the unitary operator $U(\alpha)$ parameterized by a real symmetric 2-tensor $\alpha_{ab}$,
\begin{equation}
U(\alpha) = \exp i \alpha_{ab} \Lambda^{ab}.
\end{equation}
Then, this unitary operator deforms the metric $g_{ab}$ into $g_{ab}'$ by group conjugation but leaves the determinant unchanged,
\begin{equation*}
L_{ij}(g') = U(\alpha)^{\dagger} L_{ij}(g) U(\alpha), \quad  \det g' = \det g.
\end{equation*}
If $\alpha_{ab}$ is infinitesimal, then the variation in the metric is
\begin{equation}
\delta g_{ab} = - g_{ac} \epsilon^{cd} \alpha_{db} + a\leftrightarrow b . \label{g_in_alpha}
\end{equation}

Suppose that we have an incompressible FQH state $|\Psi\rangle$ of the form (\ref{wf_decomp}) whose guiding-center metric minimizes the correlation energy (\ref{correlation_energy}). Then, we can define a deformed state
\begin{equation*}
|\Psi(\alpha)\rangle = U(\alpha) |\Psi\rangle .
\end{equation*}
We can find the generalized force by the adiabatic response associated with the variation $\alpha_{ab}$,
\begin{equation*}
F^{ab} = - \left.\frac{\partial E_G(g)}{\partial \alpha_{ab}}\right|_{\alpha=0} + \Gamma^{abcd} \dot \alpha_{cd}.
\end{equation*}
The first term vanishes because the correlation energy is minimized for the equilibrium guiding-center metric $g_{ab}$, and the second term is
\begin{align*}
\Gamma^{abcd}
&= -\hbar \,\text{Im} \langle \partial_{\alpha_{ab}} \Psi(\alpha)| \partial_{\alpha_{cd}} \Psi(\alpha)\rangle |_{\alpha = 0}\\
&= -  i \hbar\, \langle \Psi| [\Lambda^{ab}, \Lambda^{cd}] |\Psi\rangle.
\end{align*}
Dividing by the area $A$ occupied by the QH fluid, we find a 4-tensor $\eta_H^{abcd}$ which we identify as the guiding-center Hall viscosity tensor with raised indices,
\begin{equation}
 \eta^{abcd}_H=
 -\frac{1}{A}\Gamma^{abcd} = \frac{\hbar}{2\pi \ell_B^2} \frac{i}{N_{\Phi}} \langle \Psi| [\Lambda^{ab}, \Lambda^{cd}] |\Psi\rangle.
\end{equation}
In the active transformation, the deformation of the metric corresponds to the following mapping of $\bm R_i$,
\begin{align*}
R_i^a \rightarrow &\quad U(\alpha)^{\dagger} R_i^a U(\alpha) \\
&= R_i^a -i\alpha_{bc}[\Lambda^{bc},R_i^a] + \mc O(\alpha^2)\\
&= R_i^a + \epsilon^{ab} \alpha_{bc} R_i^c.
\end{align*}
Thus, we identify $\epsilon^{ac}\alpha_{cb}$ as the analog of the derivative of the displacement vector $\partial_b u^a$ in the classical elasticity theory\cite{Landau1986}. Then, the guiding-Hall viscosity tensor is
\begin{equation}
\eta_H\,^{ac}_{bd} = \eta_H^{aecf} \epsilon_{eb} \epsilon_{fd}.
\end{equation}
We can use the commutation relations of the guiding-center APD generators to expand the 4-tensor $\eta_H^{abcd}$ in terms of a symmetric 2-tensor $\eta_H^{ab}$,
\begin{subequations}
\begin{align}
\eta_H^{abcd} &=  \textstyle{\frac{1}{2}} (\epsilon^{ac} \eta_H^{bd} + \epsilon^{ad} \eta_H^{bc} + a \leftrightarrow b)\\
\eta_H^{ab} & = - \frac{\hbar}{2\pi \ell_B^2} \frac{1}{N_{\Phi}} \langle \Psi| \Lambda^{ab}|\Psi\rangle \label{gc_eta}
\end{align}
\label{gc_hv}
\end{subequations}
The quantity $\langle \Psi |\Lambda^{ab}|\Psi\rangle$ contains both super-extensive ($\propto N^2$) and extensive ($\propto N$) terms. The former contribution comes from the uniform background number-density $\nu/2\pi\ell_B^2$ ($\nu = N/N_{\Phi}$). This super-extensive term should be subtracted so that the guiding-center Hall viscosity is regularized. The extensive term does not vanish only if the electrons develop correlation. \\

Now, consider the Landau-orbit degree of freedom. After replacing $\Lambda^{ab}$ with $\tilde \Lambda^{ab}$ and $L_{ij}(g)$ with $\tilde L(\tilde g)$, the same argument works. The Landau-orbit Hall viscosity tensor is
\begin{subequations}
\begin{align}
\tilde \eta_H\,^{ac}_{bd} & =\tilde \eta_H^{aecf} \epsilon_{eb} \epsilon_{fd}\\
\tilde \eta_H^{abcd} &=  \textstyle{\frac{1}{2}} (\epsilon^{ac} \tilde\eta_H^{bd} + \epsilon^{ad}\tilde \eta_H^{bc} + a \leftrightarrow b)\\
\tilde \eta_H^{ab} & =  \frac{\hbar}{2\pi \ell_B^2} \frac{1}{N_{\Phi}} \langle \Psi| \tilde \Lambda^{ab}|\Psi\rangle \label{lo_eta}
\end{align}
\label{l_hv}
\end{subequations}
The Landau-orbit Hall viscosity does not need  regularization. The sign difference between (\ref{gc_eta}) and (\ref{lo_eta}) originates from the commutation relations of Landau-orbit and guiding center APD generators, cf.(\ref{APD_comm}). This Landau-orbit Hall viscosity exists whether or not the electrons are correlated. If the single-particle energy is of the form (\ref{simplified_single_particle_energy}), then the Landu-orbit Hall viscosity tensor can be expressed in terms of Landau-orbit spin,
\begin{equation}
\tilde \eta_H^{ab} = \frac{\hbar}{2\pi \ell_B^2} \nu \tilde s_n \tilde g^{ab}
\end{equation}
This is the Hall viscosity first discussed by Avron, Seiler and Zograf in the Galilean invariant system.\cite{Avron1995}\\

\subsection{Guiding-center spin}

In the last section, we derived the two kinds of Hall viscosity without assuming Galilean and rotational symmetry. Furthermore, there was no assumption about the shape of the QH fluid (it could take any shape as in Fig.\ref{arbitrary_edge}). In  this section, we take the shape of the QH fluid to be a ``droplet", and then we extract a quantity called the ``guiding-center spin" which is an emergent spin associated with a ``composite boson". Then, we express the guiding-center Hall viscosity in terms of the guiding-center spin.

Suppose we have a ``droplet" of the incompressible $\nu = p/q$ FQH state $|\Psi_{p/q}\rangle $ of the form (\ref{wf_decomp}) which is a condensate of ``composite bosons". Suppose its Landau-orbit metric $\tilde g_{ab}$ and guiding-center metric $g_{ab}$ take their equilibrium values.

A ``composite boson" is made of $p$ particles (which can be either fermion or boson) with $q$ flux quanta. The droplet contains $N$ ``elementary" particles so that there are $\bar N = N/p$ composite particles. The droplet is penetrated by $N_{\Phi} = q \bar N$ flux quanta.
 If we exchange two composite bosons, the state acquires a phase $\xi^p$ from the particle statistics($\xi = 1$ for the bosonic particle and $\xi = -1$ for the fermionic particle) and the Aharonov-Bohm phase $(-1)^{pq}$. The composite object is a boson, and so these two phases should cancel $\xi^p \times (-1)^{pq} = 1$.
This imposes a condition on possible combinations of the integers $p$ and $q$. The model incompressible FQH states under our consideration all satisfy this condition : bosonic Laughlin states with $p = 1$ and even $q$, fermionic  Laughlin states with $p = 1$ and even $q$, bosonic  Moore-Read state with $p = 2$ and $q = 2$, and fermionic Moore-Read state with $p = 2$ and $q = 4$

 The ``droplet" means that it is an eigenstate of the ``guiding-center rotation generator"
 \begin{subequations}
\begin{align}
L(g) &= g_{ab} \Lambda^{ab} \\
L(g)|\Psi_{p/q}\rangle &= (\textstyle{\frac{1}{2}} p q N^2 + s \bar N ) |\Psi_{p/q}\rangle. \label{guiding_center_spin}
\end{align}
\end{subequations}
where the second equation defines the rational number $s$. (In the language of the wavefunctions in the symmetric gauge, the eigenvalue of $L(g)$ is the sum, the total power of all $z_i = x_i + i y_i$ in a monomial plus $\frac{1}{2} N $.) Note that the first term is the guiding-center angular momentum from the uniform occupation profile $\bar n_m = p/q$,
\begin{equation*}
{\textstyle{\frac{1}{2}}} p q N^2 = \sum_{m= 0}^{N_{\Phi}-1} (m+ {\textstyle{\frac{1}{2}}}) \bar n_{m+1/2}.
\end{equation*}
As an analogue of the usual decomposition $J_z =  L_z +  S_z$ (the total angular momentum is the sum of orbital angular momentum and the spin), we may regard the extensive term $s \bar N$ as the spin part of the total angular momentum from $\bar N$ composite bosons. We call $s$ the ``guiding-center spin".

Let's calculate $s$ for Laughlin $1/3$ state as an example. Since the Laughlin state is a Jack polynomial\cite{Bernevig2008} with the proper normalization factors (cf. Sec.\ref{sec:mapping}), its guiding-center angular momentum can be calculated from the ``root occupation profile" $\{n^0_{m +1/2} : m\in \mb Z_+\} = \{n_{1/2}^0 , n_{3/2}^0, n_{5/2}^0, n_{7/2}^0, \dots\} = \{1,0,0,1,0,\dots\}$. Its root occupation profile is a repetition of the pattern $(1,0,0)$. The guiding-center angular momentum is then
\begin{equation*}
\sum_{m = 0}^{N_{\Phi}-1} (m + {\textstyle{\frac{1}{2}}}) n_{m+1/2}^0 = { \textstyle{\frac{3}{2}}} N^2 - \bar N
\end{equation*}
Comparing this with (\ref{guiding_center_spin}), we deduce $s = -1$ for the Laughlin $1/3$ state. The guiding-center spins of Laughlin $1/q$ state for $q = 2,3,4$ and Moore-Read $2/q$ state for $q = 2,4$ are listed in the Table.\ref{expected}. Note that the guiding-center spin vanishes for uncorrelated uniform states.

\begin{table}
\resizebox{4cm}{1.25cm}{
\begin{tabular}{*{6}{c}}
\hline\hline
$\nu $& $\frac{1}{2}$ & $\frac{1}{3}$ & $\frac{1}{4}$& $\frac{2}{2}$ & $\frac{2}{4}$\\\hline
$s$ &$- \frac{1}{2}$ & $-1$ & $-\frac{3}{2}$ &$-1$ &$-2$\\
$\frac{-s}{q}$& $\frac{1}{4}$ & $\frac{1}{3}$ & $\frac{3}{8} $& $\frac{1}{2}$ & $\frac{1}{2}$\\
\hline\hline
\end{tabular}}
\caption{Guiding center spin $s$.}
\label{expected}
\end{table}

From the fact that $|\Psi_{p/q} \rangle$ is the eigenstate of $L(g)$, we can calculate its expectation value of the \textit{regularized}  guiding-center APD generator $\delta \Lambda^{ab}$,
\begin{equation}
\langle \Psi_{p/q} | \delta \Lambda^{ab}  |\Psi_{p/q}\rangle
= \textstyle{\frac{1}{2}} g^{ab}  s \bar N.
\end{equation}
Inserting this into  (\ref{gc_eta}), we obtain the \textit{regularized} guiding-center Hall viscosity tensor,
\begin{equation}
\eta_H^{ab} = -\frac{\hbar}{4\pi \ell_B^2} \frac{s}{q} \,g^{ab}.\label{rot_hall}
\end{equation}
In general, the guiding-center metric may depend on the spatial coordinates on the length scale much larger than $\ell_B$ while the guiding-center spin remains quantized,
\begin{equation}
\eta_H^{ab}(\bm r) = -\frac{\hbar}{4\pi \ell_B^2} \frac{s}{q} \,g^{ab}(\bm r).
\end{equation}

From (\ref{main}), we obtain the expression of the intrinsic dipole moment per a line element $dL^a$ in terms of the guiding-center spin and the number of flux quanta in a composite boson,
\begin{equation}
B d p^a =   \frac{\hbar}{4\pi \ell_B^2} \frac{s}{q} \,g^{ab} \epsilon_{bc} d L^c.\label{gc_spin_and_dipole}
\end{equation}
 For a line element $dL^x$,
\begin{equation}
B dp^y = -\frac{\hbar}{4\pi \ell_B^2} \frac{s}{q} g^{yy} d L^x.
\end{equation}
The expected intrinsic dipole moments $dp^y$ are listed in Table.\ref{expected} in the unit $e/4\pi$ ($g_{ab} = \delta_{ab}$).
This will be verified numerically  in Sec.\ref{sec:jack} and Sec.\ref{sec:es_dipole}. \\

We recover the Hall viscosity discussed in other works\cite{Read2011, Bradlyn2012} if we impose inessential rotational invariance $g_{ab} = \tilde g_{ab}$. In this case, the usual angular momentum $L_z$ is a good quantum number.
\begin{equation*}
L_z = g_{ab} (\tilde \Lambda^{ab} - \Lambda^{ab}),
\quad \delta L_z =  g_{ab} (\tilde \Lambda^{ab} - \delta \Lambda^{ab}),
\end{equation*}
where $\delta L_z$ is the \textit{regularized} angular momentum subtracting the contribution from the uniform density. The expectation value $\delta L_z$ for the model states $|\Psi_{p/q}\rangle$ divided by the number of composite bosons $\bar N$ gives
\begin{equation*}
\bar N^{-1}\langle \Psi_{p/q} |\delta L_z|\Psi_{p/q}\rangle = p\tilde s - s ,
\end{equation*}
which is the total spin per composite boson. For $1/q$ Laughlin states, the guiding-center spin is $s = \textstyle{\frac{1}{2}}(1-q)$, and the total spin  per composite boson is $\textstyle{\frac{1}{2}} q$. This coincides with what was called ``orbital spin"  by Wen and Zee\cite{WenZee1992}, and later by Read and Rezayi\cite{Read2011}. In such rotational invariant system, the sum of the Landau-orbit and guiding-center Hall viscosities becomes
\begin{equation*}
\tilde \eta^{ab}_H +\eta^{ab}_H = \frac{\hbar}{4\pi \ell_B^2} \left(\nu \tilde s - \frac{s}{q}\right) g^{ab}.
\end{equation*}
This is the Hall viscosity  discussed by Read and Rezayi\cite{Read2011}. Note this is valid only in the rotational invariant system, and it misses the separation of two types of Hall viscosity.

%% file: jack.tex
\subsection{Mapping Jack polynomials to wavefunctions}
\label{sec:mapping}
A model bosonic quantum Hall state with $N$ particles at the filling $\nu = p/q$
is described by a symmetric Jack polynomial \cite{Bernevig2008}
\begin{equation*}
J_{\lambda_0(p,q)}^{\alpha(p,q)}(z_1, z_2,\dots z_N),
\end{equation*}
which is labeled by one negative rational parameter $\alpha(p,q)$ and a
root ``admissible partition" $\lambda_0(p,q)$  with length $\ell_{\lambda_0}\le N$.
$\alpha(p,q) = -(p+1)/(q-1)$ where $p+1$ and $q-1$ are relatively prime.
Given $p$ and $q$, a partition $\lambda$ is admissible if the partition, when
translated into a set of occupation numbers, satisfies a generalized exclusion:
there are no more than $p$ particles in $q$ consecutive orbitals. A Jack
polynomial for a model fermionic quantum Hall state at the filling $\nu = p/(p + q)$
is obtained from the symmetric Jack $J_{\lambda_0}^{\alpha}$ with by multiplying a
Vandermonde factor \cite{Bernevig2009}. For $p = 1$, the Jack polynomial corresponds
to Laughlin wavefunctions, and for $p = 2$, it corresponds to Moore-Read wavefunctions.
A monomial $m_{\lambda}$ labeled by a partition $\lambda$ for $N$ particles is defined as
 \begin{equation*}
m_{\lambda}(z_1,\dots ,z_N) \equiv \sum_{\tau \in S_N} \prod_{j = 1}^N z_{\tau (j)}^{\lambda(j)},
\end{equation*}
where $S_N$ is all permutations of $\{1,\dots ,N\}$. A bosonic (fermionic) Jack
parameterized by a root admissible partition $\lambda_0(p,q)$ is spanned only
by monomials $m_{\lambda}$ (slater determinants $sl_{\lambda}$ ) with partitions
$\lambda$ that are obtainable by ``squeezing" the root admissible partition
$\lambda_0(p,q)$: one squeezing operation corresponds to changing
$\lambda(j)\rightarrow \lambda(j) - 1$ and $\lambda(k) \rightarrow \lambda(k) +1$
for a pair $(j,k)$ such that $j < k \in \{1,\dots \ell_{\lambda}\}$.
That is, a symmetric Jack can be written as
\begin{equation}
J_{\lambda_0}^{\alpha} = \sum_{\lambda \le \lambda_0} a_{\lambda_0,\lambda}(\alpha) m_{\lambda}.
\end{equation}
 There is a recursion relation for the rational expansion coefficients
$a_{\lambda_0,\lambda}(\alpha)$ with $a_{\lambda_0,\lambda_0} (\alpha)= 1$ \cite{Bernevig2009}.
These recursion relations allow us to generate model quantum Hall states with a
large number of particles: for this report, we used Jacks with 14 and 15 particles
for $\nu = 1/2$ bosonic Laughlin state and $\nu = 1/3$ fermionic Laughlin
state\cite{Laughlin1983}. For $\nu = 1/4$ bosonic Laughlin state, we used a
Jack with 11 particles. We used Jacks with 18 and 20 particles for $\nu = 2/2$
bosonic Moore-Read state and $\nu = 2/4$ fermionic Moore-Read state.
The MR states we used are in topologically trivial sectors: the MR 2/2 state has the root occupation pattern 2020...202 and the MR 2/4 state has the root occupation pattern 11001100...110011. \cite{Li2008}

 A Jack with variables $\{z_1,z_2,\dots ,z_N\}$ knows only about
the ``clustering property"\cite{Bernevig2008}, and it becomes physical only
after we map monomials $m_{\lambda}$ spanning the Jack into states in a
Landau Level depending on the geometry where the Hall fluid is placed on,
such as a cylinder, sphere or plane.  We map each $z_j^{m}$ for $m\in \mathbb{Z}^+$
in the monomial into a single particle wavefunction :
\begin{equation}
z_j^{m}   \rightarrow  w(m)\langle \bm r_j|m\rangle \label{state_mapping},
\end{equation}
where $|m\rangle$ is a geometry-dependent normalized single particle wavefunction
with quantum number $m$ in the lowest Landau Level (we may work within other Landau
level) and $w(m)$ is the inverse of the geometry-dependent normalization factor. Then, the
monomial $m_{\lambda}$ maps to a normalized $N$-particle wavefunction $|\Psi_{\lambda}\rangle$:
\begin{equation}
m_{\lambda} \rightarrow \langle\{\bm r_j\}|\Psi_{\lambda}\rangle =\frac{\prod_{j=1}^N
w(\lambda(j))}{\sqrt{N!}}\sum_{\tau \in S_N} \prod_{j = 1}^N  \langle \bm r_{\tau(j)}|\lambda(j)\rangle.
\end{equation}
Finally, the Jack polynomial maps to a physical model quantum
Hall state $|\Psi_{\lambda_0}^{\alpha}\rangle$ (without overall normalization):
\begin{equation}
J_{\lambda_0}^{\alpha}\rightarrow\langle\{\bm r_j\} |\Psi_{\lambda_0}^{\alpha}\rangle=
\sum_{\lambda \le \lambda_0} a_{\lambda_0,\lambda} (\alpha) \langle\{\bm r_j\}|\Psi_{\lambda}\rangle \label{jack}.
\end{equation}

\subsection{Fermi momenta}
\label{sec:Fermi}
 We consider cylinders periodic with circumferences of different lengths $L$ along
the edge direction $\bm{\hat x}$ and infinite in the direction $\bm {\hat y}$, i.e. we
use the Landau gauge. In the lowest Landau level, the normalized single-particle states
$\phi_k(\bm r)$ are labeled by the momentum $k$ along $\hat {\bm x}$ direction:
\begin{align*}
\langle \bm r|k\rangle &=\phi_{k}(\bm r) =
\frac{ e^{-(k\ell_B)^2/2}}{(\pi)^{1/4}(\ell_B L)^{1/2}}z^k e^{-(y/\ell_B)^2/2} \nonumber\\
& z = e^{i(x-iy)}.
\end{align*}
Let us write the wave-vector as $ k = 2\pi m /L$. If  the underlying constituent particles of a
quantum Hall state are bosons, then the allowed values of $m$ are $\{m \in \mathbb Z\}$
and  if they are fermions, $ \{m \in \mathbb Z+1/2\}$. This is because at zero temperature,
the chemical potential is located at a single-particle energy level for bosons, and it is located
half way between two consecutive energy levels for fermions.

Now, we would like to map Jack polynomials into  physical states as described in the previous section.
However, the mapping (\ref{state_mapping}) is not uniquely defined:
consider the bosonic case. For $m\in \mathbb Z_+$, if the term $z_i^m$ in a monomial
is  mapped to the single-particle state with definite momentum $k = (2\pi/L) m$, then
another mapping that maps $z^m$  to a state with momentum $k = (2\pi/L) (m+M)$
is also possible for any fixed integer $M$. This arbitrariness is removed when we choose
one of the Fermi momenta to be $k = 0$, and the first occupied state to have a momentum
$k = (2\pi/L) m_0$. With this mapping, the inverse of the normalization factor is
\begin{equation*}
w(m) = (\pi)^{1/4}(\ell_B L)^{1/2} \exp\left[\frac{1}{2}\left(\frac{2\pi \ell_B (m+m_0)}{L}\right)^2\right].
\end{equation*}

We want to determine the quantum number $m_0$ for the first occupied state. For instance, consider a Laughlin $\nu = 1/q$  state, the number $m_0$ is
fixed by its chiral boson edge theory : its first non-zero occupation occurs at the
momentum $k = \pi q/L$. This can be seen by Fourier-transforming the electron
Green's function in the chiral boson theory \cite{Wen1990}
\begin{equation*}
G(x-y) \propto \left( \sin\left(\pi (x-y-i\eta)/L\right)\right)^{-q},\quad \eta\rightarrow 0^+.
\end{equation*}
From this, we can obtain the expectation value of the occupation number operator
of the Laughlin state
\begin{equation*}
\langle n_{m}\rangle \propto \frac{(m+q/2-1)!}{(q-1)!(m-q/2)!}.
\end{equation*}
For $\nu = 1/2$, the occupation behaves as $\langle n_m\rangle \propto m$.
The first occupied
state corresponds to $m = 1$. Thus, a factor $z_i^0$ in a monomial $m_{\lambda}$ should be
mapped to the single-particle state with  $k = (2\pi/L)m_0=2\pi/L$. For $\nu = 1/3$,
$\langle n_m\rangle \propto (m+1/2)(m-1/2)$. The first occupied state corresponds to $m = 3/2$,
so we should map $z_i^0$ to the single-particle state with $k = (2\pi/L)m_0=(2\pi/L) (3/2)$.
For $\nu = 1/4$, $\langle n_m \rangle \propto (m+1)m(m-1)$. In general, we have
$m_0 = q/2$ for  $\nu = 1/q$ Laughlin state.

\begin{figure}[htb]
\includegraphics[width=1\linewidth]{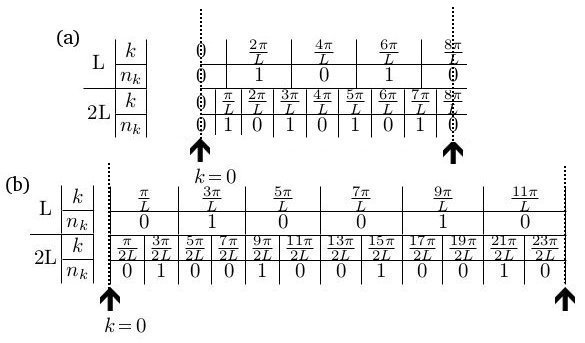}
\caption{$N=2$ and $N = 4$ root occupations for (a) $\nu = 1/2$ and
(b) $\nu = 1/3$ Laughlin states with two different circumferences $L$ and $2L$.
The two bold arrows denote two Fermi momenta. For $\nu =1/2$, there
 are $4 = 2\cdot 2$ and $8 = 2\cdot 4$ states  respectively between the
two Fermi momenta. For $\nu =1/3$, there are $6 = 3\cdot 2$ and
$12 = 3\cdot 4$ states respectively. }
\label{root_occ}
\end{figure}

 This implies that when a $N = p \bar N$ particle quantum Hall state with
the filling factor $\nu = p/q$ is put on a cylinder, then between the two
Fermi momenta there are $q\bar N$ orbitals. For example, for $\nu =1/2$
and $\nu = 1/3$ Laughlin states, we have the following (Fig.\ref{root_occ})
root momentum occupations (i.e. the occupations of the state corresponding
to the monomial $m_{\lambda_0}$) for two circumferences $L$ and $2L$.

In order to have the two edges not interact with each other, we need
to take a limit $N\rightarrow \infty$ first, and then take $L\rightarrow \infty$.
In practice, we can have only finite $N$, and this restricts the largest available
$L$ for the fixed $N$. If $L$ increased further than this value, then the Jack
 polynomial becomes a wavefunction of Calogero-Sutherland model with its
two edges interacting strongly \cite{Haldane1994}. If $L$ is too small, then
the Jack becomes a charge-density-wave state. For fixed $L$, the occupation numbers
converge to some limits as the number of particles $N$ increases.

\subsection{Occupation number}
\label{sec:jack_occ}

At zero temperature, $N$ non-interacting electrons in an
IQH fluid fill up the states from $m = 1/2$ to $m=N/2$ with
$\langle n_m \rangle = 1$. In the case of FQH fluid with filling
factor $\nu$, not only the range of momenta of occupied states
changes so as to satisfy $\langle n_m\rangle \approx \nu$ but
also $\langle n_m\rangle$ deviates from $\nu$ appreciably
near the Fermi  momenta. This variation of $\langle n_m\rangle$
gives rise to the intrinsic dipole moment. We analyze the
occupation numbers.

Given a ground state wavefunction $|\Psi_{\lambda_0}^{\alpha}\rangle$
that we obtain from a Jack polynomial $J_{\lambda_0}^{\alpha}$ by a
mapping described in the preceding section, we can calculate the occupation number (i.e. the
expectation value of the occupation number operator $n_m $) for each
momentum $k = 2\pi m/L$.  We evaluate
\begin{equation*}
\langle n_{m}\rangle_0\equiv \frac{\langle\Psi_{\lambda_0}^{\alpha}
| n_m|\Psi_{\lambda_0}^{\alpha}\rangle}{\langle\Psi_{\lambda_0}^{\alpha}
|\Psi_{\lambda_0}^{\alpha}\rangle}= \frac{\sum_{\lambda \le \lambda_0}
a_{\lambda_0,\lambda}(\alpha)^2\langle\Psi_{\lambda} | n_m
|\Psi_{\lambda}\rangle}{\sum_{\lambda \le \lambda_0}
a_{\lambda_0,\lambda}(\alpha)^2\langle\Psi_{\lambda} |\Psi_{\lambda}\rangle},
\end{equation*}
where $\lambda\le \lambda_0$ means that the sum is over all
partitions that are obtainable by squeezing from the root partition $\lambda_0$.
Note that $\langle n_0 \rangle_0 = 0$ for $\nu = 1/2$ Laughlin state,
$\langle n_{1/2}\rangle_0 = 0$ for $\nu =1/3$ Laughlin state, and so on.

\begin{figure}[htb]
\includegraphics[width=0.85\linewidth,page =3]{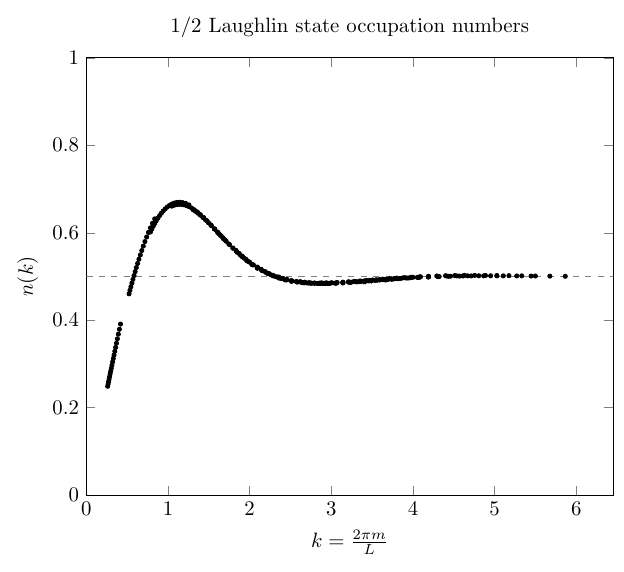}
\caption{$\nu = 1/2$ Laughlin state density profile : $(\times)$  for $N = 14$ and $(\bullet)$ for $N = 15$. It plots data obtained with different
$L = 15$ to $24$ with increments by $0.5$ (in units of $\ell_B$).
The horizontal line is 1/2. The linear fit of $\log(k = 2\pi/L)$ versus
$\log n(k)$ gives $ \log n(k) = 0.963 \log k -0.010$ with the norm of residues $0.003$ }
\label{l2_profile}
\end{figure}

\begin{figure}[htb]
\includegraphics[width=0.85\linewidth, page = 4]{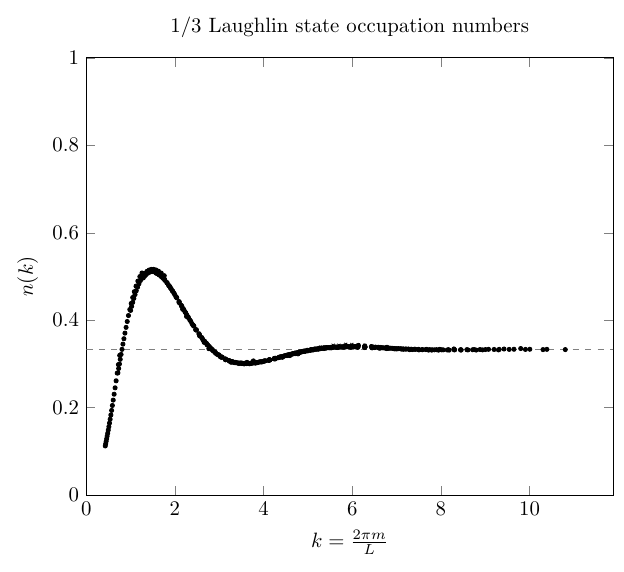}
\caption{$\nu = 1/3$ Laughlin state density profile : $(\times)$  for $N = 14$ and $(\bullet)$ for $N = 15$. $L = $ 12.5 to 22 with increments by 0.5. The horizontal line is 1/3.
The linear fit of $\log(k = 3\pi/L)$ versus $\log n(k)$ gives $\log n(k) =
1.853 \log k -0.609$ with the norm of residues $0.017$.
}
\label{l3_profile}
\end{figure}

\begin{figure}[htb]
\includegraphics[width=0.85\linewidth, page = 5]{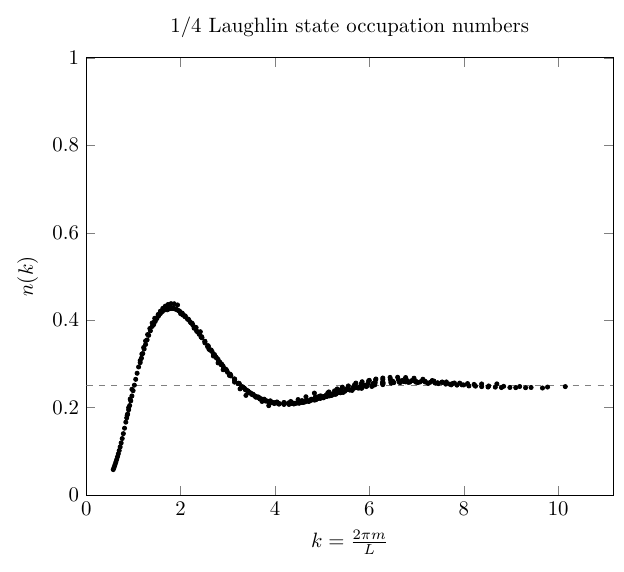}
\caption{$\nu = 1/4$ Laughlin state density profile :  $(\bullet)$
 for $N = 11$. $L = $ 12.5 to 22 with increments by 0.5. The horizontal line is 1/4.
The linear fit of $\log(k = 4\pi/L)$ versus $\log n(k)$ gives $\log n(k) = 2.722\log k -0.530 $
with the norm of residues $0.028$
}
\label{l4_profile}
\end{figure}

\begin{figure}[htb]
\includegraphics[width=0.85\linewidth, page = 6]{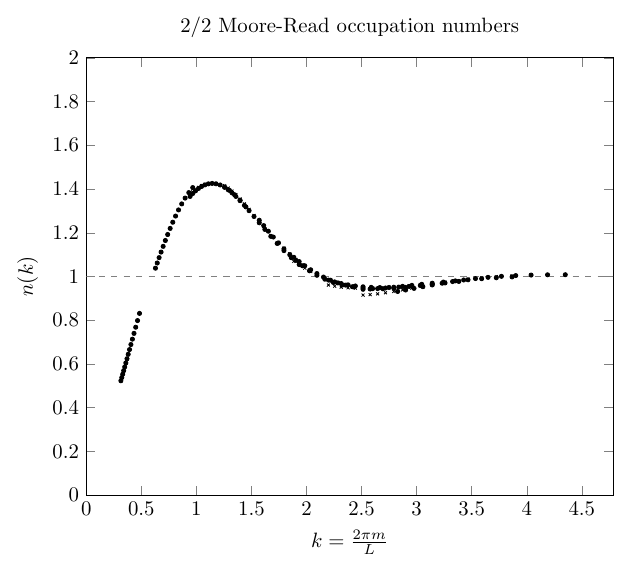}
\caption{$\nu = 2/2$ Moore-Read state density profile: $(\times)$ for $N = 18$ and $(\bullet)$  for $N = 20$. $L = $ 13 to 20 with increments by 0.5. The horizontal line is 1. The linear fit of $\log(k = 2\pi/L)$
versus $\log n(k)$ gives $\log n(k) = 1.076 \log k +0.598$ with the norm of residues $0.002$.
}
\label{mr1_profile}
\end{figure}

\begin{figure}[htb]
\includegraphics[width=0.85\linewidth,  page = 7]{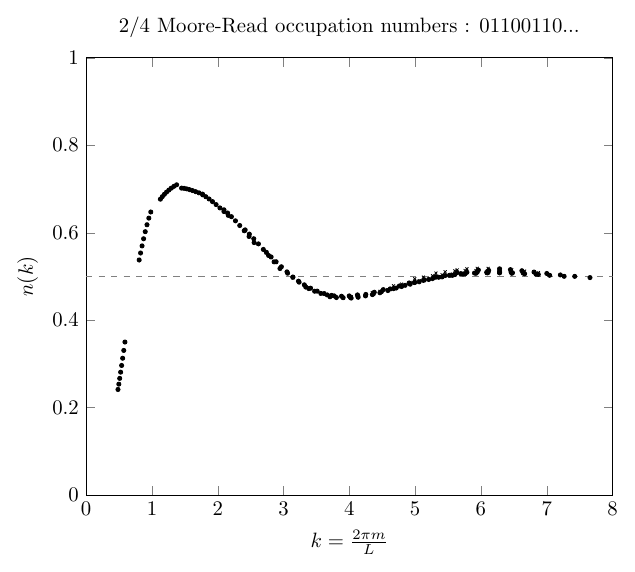}
\caption{$\nu = 2/4$ Moore-Read state density profile:$(\times)$ for $N = 18$ and $(\bullet)$  for $N = 20$. $L = $ 16 to 19.5 with increments by 0.5. The horizontal line is 1/2. The linear fit of $\log(k = 3\pi/L)$ versus $\log n(k)$  gives $\log n(k) = 1.879 \log k - 0.054$ with the norm of residues $0.002$.
}
\label{mr2_profile}
\end{figure}

The  occupation numbers are calculated for the model wavefunctions
and are plotted in Fig.\ref{l2_profile}, \ref{l3_profile}, \ref{l4_profile}, \ref{mr1_profile}
and \ref{mr2_profile}. The first two plots which are occupations of Laughlin 1/2 and 1/3
states have total numbers of particles $N =$ 14 and 15. The next plot is that of
Laughlin 1/4 state with a total number of particles $N = 11$.
The last two plots are occupations of Moore-Read 2/2 and 2/4 states have total
numbers of particles $N=$ 18 and 20.  These plots show only half of the occupation
profile because the other half can be obtained by mirror symmetry.
The  occupation numbers are plotted as a function of the momentum $k$ rather than the quantum number $m$,
\begin{equation*}
n(k) = \langle n_m\rangle_0, \quad k = \textstyle{\frac{2\pi}{L}} m.
\end{equation*}

The figures contain data for several values of $L$ on the same plot. Each occupation plot seems to follow a smooth profile that might appear in the
limit $L\rightarrow \infty$.
This observation allows us to observe how well these wavefunctions of
finite numbers of particles agree with the behavior of $n(k)$ near $k= 0$
described by the chiral boson theory. In each occupation profile plot,
we calculate the linear fit  of $\log n(k)$ versus $\log k$ with those
momenta $k = (2\pi/L)m_0$ , the first non-vanishing occupation numbers.
We observe that they are quite linear, and their linear fit coefficient is the
exponent $r$ in $n(k) \propto k^r$ as $k \rightarrow 0$. For each Laughlin
$1/2,\,1/3,\,1/4$ state, the exponent is calculated to be $0.963,\, 1.853,\, 2.722$
respectively, while
 the expected exponents are 1, 2 and  3. For each Moore-Read 2/2 and 2/4 state,
the exponent is calculated to be 1.076 and 1.879 while the expected exponents are 1 and 2.

\begin{figure}[htb]
\includegraphics[width=0.85\linewidth, page = 10]{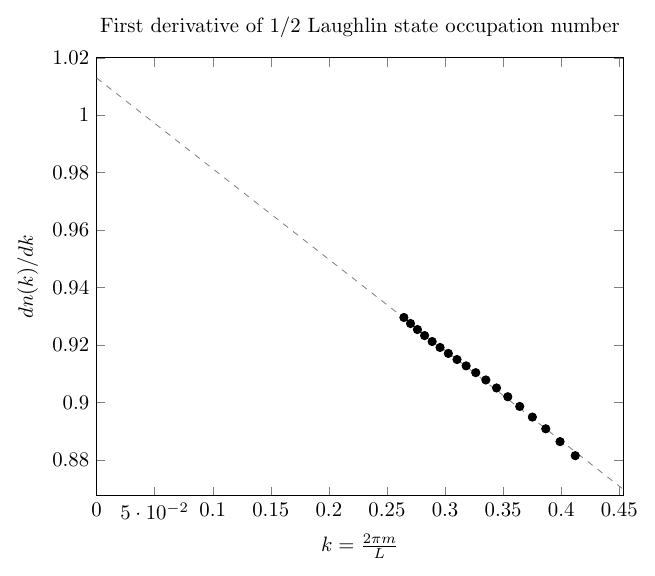}
\caption{ $n'(k)$ of $\nu = 1/2$ Laughlin state: we plot occupation
numbers with $k = \frac{2\pi}{L} $. The linear fit  (dashed line)
gives $n'(k) = -0.316 k + 1.013$ with norm of residuals $0.0024$.
}
\label{l2_dn}
\end{figure}

\begin{figure}[htb]
\includegraphics[width=0.85\linewidth, page = 11]{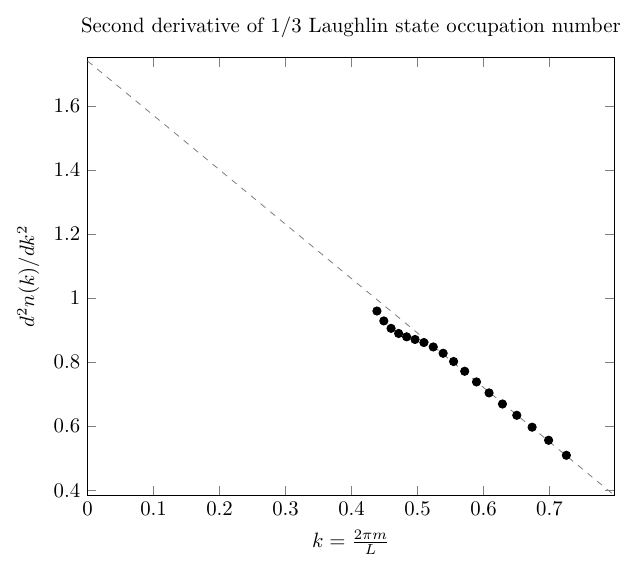}
\caption{ $n''(k)$ of $\nu = 1/3$ Laughlin state: we plot occupation
numbers with $k = \frac{3\pi}{L} $. The linear fit (dashed line) gives
$n''(k) = -1.698 k + 1.740$ with norm of residues $0.0069$.}
\label{l3_ddn}
\end{figure}

Moreover, if we assume the exponents from the chiral boson theory and
accept the form of occupation number $n(k) = A \, k^{q-1}+\dots$ near $k=0$,
then we can calculate the numerical factor $A$ which is inaccessible in the field theory.
Defining derivatives of $n(k)$ by finite difference, we obtain $n'(k)$ for Laughlin $1/2$ state
and $n''(k)$ for Laughlin $1/3$ state. These are plotted in Fig. \ref{l2_dn} and \ref{l3_ddn},
respectively. We find that near $k=0$, $n(k) = 1.013k + \mathcal O(k^2)$ for Laughlin 1/2
state, and $n(k) = 0.870k^2+\mathcal O(k^3)$ for Laughlin 1/3 state. We notice that these
numerical factors $A$ might become rational numbers such as 1 and 7/8 respectively in the
thermodynamic limit.



\subsection{Luttinger's theorem}
\label{sec:luttinger}
Given the occupation numbers, we can also verify the that they
satisfy the Luttinger sum rule\cite{Luttinger1960}.  For a one dimensional
system with ``Fermi surface'' singularities in the occupation numbers
$n(k)$ at ``Fermi points'' $k_i$, this states that
\begin{equation}
\frac{N}{L} =  \int \frac{dk}{2\pi}\, n(k) =   \int \frac{dk}{2\pi} \, n_0(k),
\end{equation}
 where in a Luttinger liquid (the  1D  analog of a Fermi liquid),
$n_0(k)$ is a  integer topological index  that is constant in
regions $k_i < k< k_{i+1}$ and counts the number of occupied bands
below the Fermi level with  momentum or Bloch index $k$.
(From a ``modern'' viewpoint, the Luttinger theorem
is an early example of the identification of a topological index
 $n_0(k)$ that remains invariant as the actual $n(k)$ is continuously
modified by the interactions in the Fermi liquid that conserve
the existence of the  singularity at the Fermi  surface.)
  In the fractional quantum Hall effect in the $L\rightarrow \infty$
limit of the
cylinder geometry,  this
generalizes to  $n_0(k)$ = $\nu (k)$, the filling factor  in the
region   $ k_i\ell^2_B < y < k_{i+1}\ell_B^2$.

The applicability
of the Luttinger theorem to the fractional quantum Hall
fluid\cite{Haldane1993, varjs} is
immediately visible in the Jack polynomial description:
the ``root'' configuration of, \textit{e.g.},
 the $\nu$ = $\frac{1}{3}$ Laughlin state is
$\ldots000|010010010\ldots 010010010|000\ldots$ with a mean occupation of
$\nu$ = $\frac{1}{3}$ between the  Fermi points  marked as ``$|$''.
This is a uniform filling  $\nu$  in the thermodynamic limit.
The ``squeezing'' of pairs of ``1'''s together in the full Jack
configuration preserves this mean filling in the interior of strips
much wider than $\ell_B$, creating the dipoles near the Fermi
points, and preserving the Luttinger sum rule.
The Luttinger sum rule is the integral  form of the differential
relation $dN = (L/2\pi)\sum_i \Delta \nu_i dk_i$, where
$\Delta \nu_i$ = $\nu (k=k_i^+)- \nu(k=k_i^-)$ is the chiral anomaly of the Fermi point\cite{Haldane1993}.\\

\begin{figure}[htb]
\includegraphics[width=0.85\linewidth, page = 12]{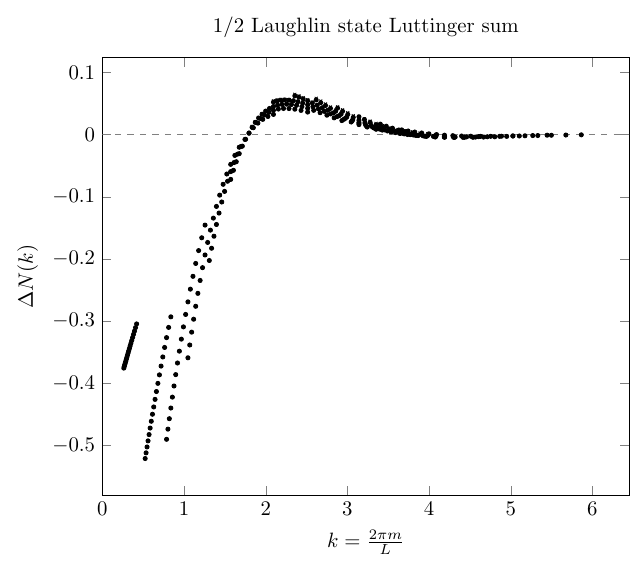}
\caption{$\Delta N(k)$ for $\nu = 1/2$ Laughlin state : $(\times)$ for $N = 14$  and $(\bullet)$ for $N = 15$. It plots data obtained with  $L =$ 15 to 24 with increments by 0.5.  }
\label{l2_luttinger}
\end{figure}

\begin{figure}[htb]
\includegraphics[width=0.85\linewidth, page = 13]{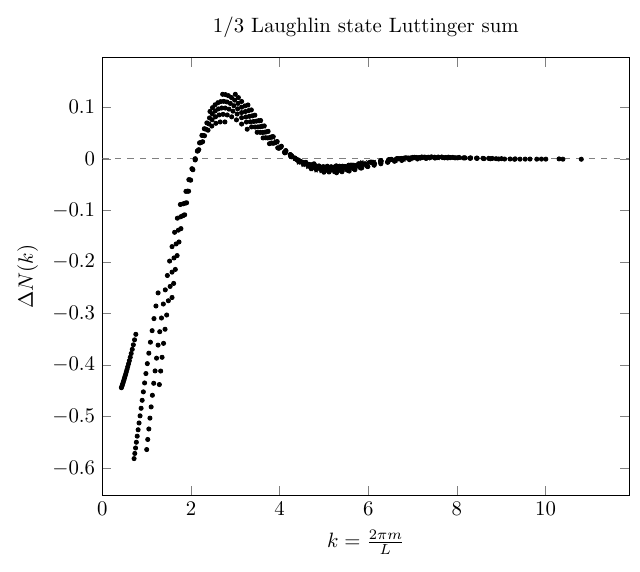}
\caption{ $\Delta N(k)$ for $\nu = 1/3$ Laughlin state :  $(\times)$ for $N = 14$  and $(\bullet)$ for $N = 15$. $L =$ 12.5 to 22  with increments by 0.5.
}
\label{l3_luttinger}
\end{figure}

\begin{figure}[htb]
\includegraphics[width=0.85\linewidth, page = 14]{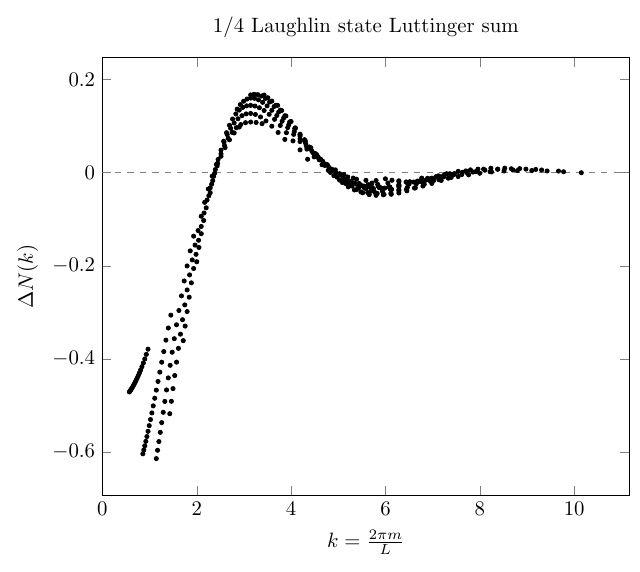}
\caption{ $\Delta N(k)$  for $\nu = 1/4$ Laughlin state:  $(\bullet)$ for $N = 11$.  $L = 12.5$  to $22$  with increments by $0.5$.
}
\label{l4_luttinger}
\end{figure}

\begin{figure}[htb]
\includegraphics[width=0.85\linewidth, page = 15]{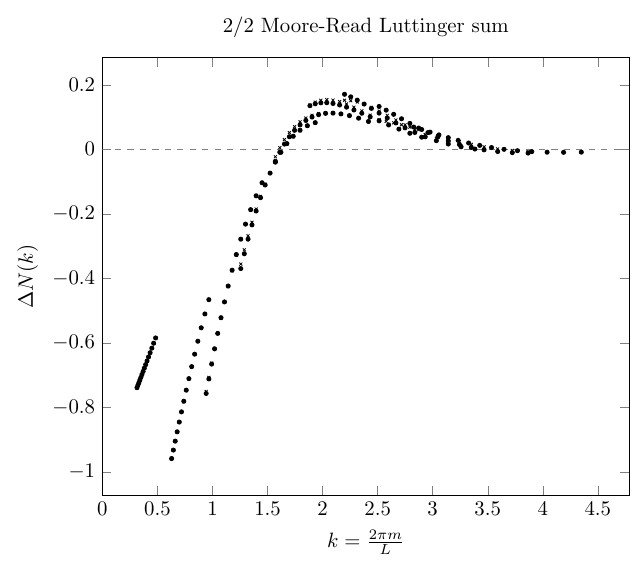}
\caption{ $\Delta N(k)$  for $\nu = 2/2$ Moore-Read state  : $(\times)$ for $N = 18$ and $(\bullet)$  for $N = 20$. $L = 13$  to $20$ with increments by 0.5.
}
\label{mr1_luttinger}
\end{figure}

\begin{figure}[htb]
\includegraphics[width=0.85\linewidth, page = 16]{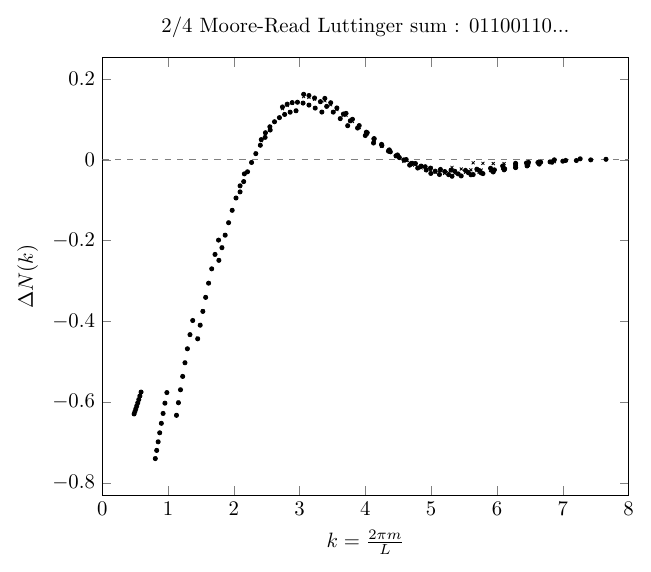}
\caption{ $\Delta N(k)$  for $\nu = 2/4$ Moore-Read state : $(\times)$ for $N = 18$ and $(\bullet)$  for $N = 20$. $L = 16$  to $19.5$ with increments by 0.5.
}
\label{mr2_luttinger}
\end{figure}

We define the function $\Delta N(k)$ which is the integration  of the difference between the actual occupation number  and the uniform occupation number  from $0$ to $k$ is
\begin{equation*}
\frac{\Delta N(k)}{L} = \int_0^k \frac{d k'}{2\pi} ( n(k') - \nu).
\end{equation*}
For finite $L$, the integration is approximated by the sum,
\begin{align}
\Delta N\left(k\right) &= \sum_{m' = 0 \text{ or } 1/2}^{m-1}\langle n_{m'}\rangle_0+\textstyle{\frac{1}{2}}\langle n_m\rangle_0-\nu m,
\end{align}
where the summation is over integers $m' = 0,1,2,\dots, m-1$  for a bosonic state, and it is over half-integers $m' =\textstyle{\frac{1}{2}},\textstyle{\frac{3}{2}},\textstyle{\frac{5}{2}},\dots, m-1$ for a fermionic state.
If the Luttinger's theorem holds this should vanish as $k$ gets larger.
Because we are limited by the finite size, we calculate $\Delta N(k)$ only
up to the center of the fluid. $\Delta N(k)$ is plotted against $k$. Each
plot includes data from a range of circumferences $L$.
See Fig . \ref{l2_luttinger}, \ref{l3_luttinger}, \ref{l4_luttinger}, \ref{mr1_luttinger}
and \ref{mr2_luttinger}. We observe Luttinger's theorem indeed
holds in presence of interactions among particles.


\subsection{Intrinsic dipole moment}
\label{sec:dipole}

\begin{figure}[htb]
\includegraphics[width=0.9\linewidth, page = 18]{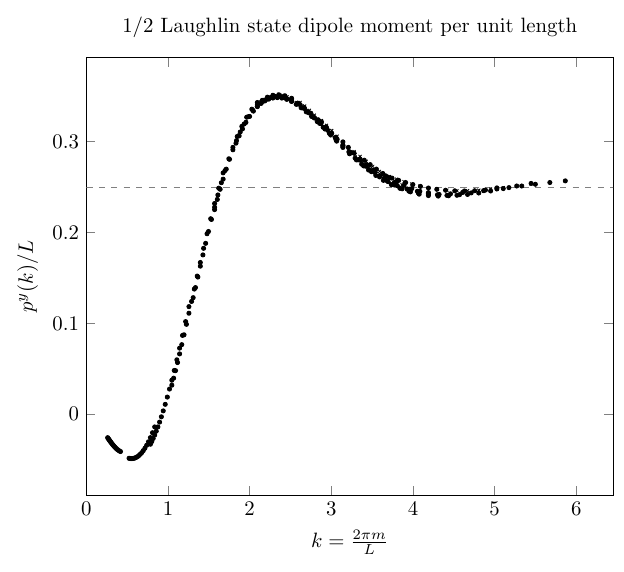}
\caption{ $p(k)/L$ in units of $-e/4\pi$ for $\nu = 1/2$ Laughlin state. Calculated from Fig.\ref{l2_profile}. $(\times)$ for $N = 14$  and $(\bullet)$ for $N = 15$. It plots data obtained with  $L =$ 15 to 24 with increments by 0.5.  It converges to $1/4$.}
\label{l2_dipole}
\end{figure}

\begin{figure}[htb]
\includegraphics[width=0.9\linewidth, page = 19]{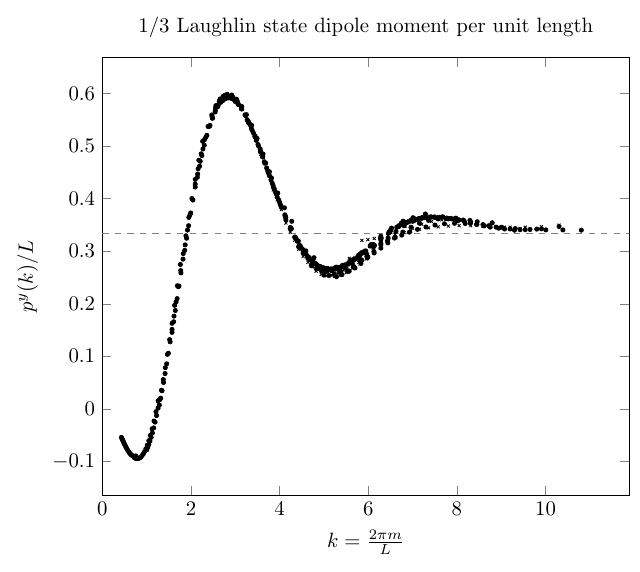}
\caption{$p(k)/L$ in units of $-e/4\pi$ for $\nu = 1/3$ Laughlin state. $(\times)$ for $N = 14$  and $(\bullet)$ for $N = 15$. $L =$ 12.5 to 22  with increments by 0.5.
It converges to $1/3$.}
\label{l3_dipole}
\end{figure}

\begin{figure}[htb]
\includegraphics[width=0.9\linewidth, page = 20]{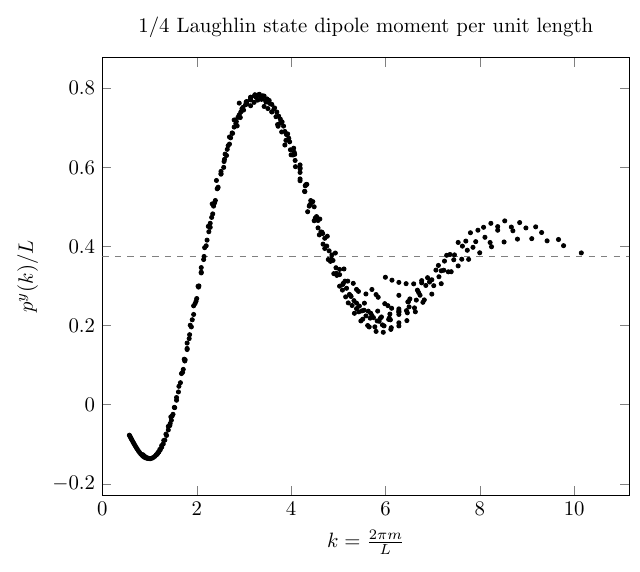}
\caption{$p(k)/L$ in units of $-e/4\pi$ for $\nu = 1/4$ Laughlin state. $(\bullet)$ for $N = 11$.  $L = 12.5$  to $22$  with increments by $0.5$.
It converges to $3/8$.}
\label{l4_dipole}
\end{figure}

\begin{figure}[htb]
\includegraphics[width=0.9\linewidth, page =21]{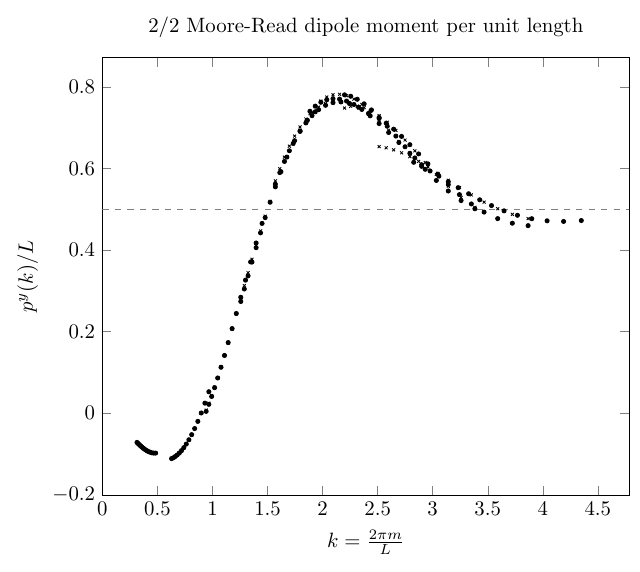}
\caption{$p(k)/L$ in units of $-e/4\pi$ for $\nu = 2/2$ Moore-Read state.$(\times)$ for $N = 18$ and $(\bullet)$  for $N = 20$. $L = 13$  to $20$ with increments by 0.5. It converges to $1/2$.}
\label{mr1_dipole}
\end{figure}

\begin{figure}[htb]
\includegraphics[width=0.9\linewidth, page = 22]{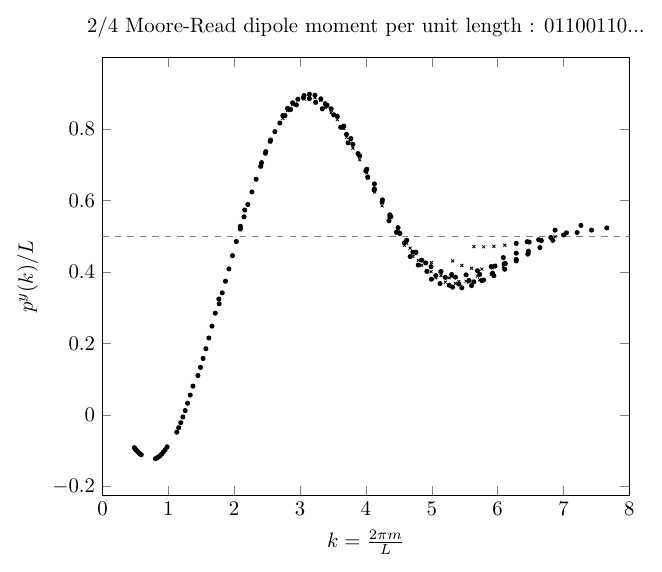}
\caption{$p(k)/L$ in units of $-e/4\pi$ for $\nu = 2/4$ Moore-Read state. $(\times)$ for $N = 18$ and $(\bullet)$  for $N = 20$. $L = 16$  to $19.5$ with increments by 0.5. It converges to $1/2$.}
\label{mr2_dipole}
\end{figure}

Here, we calculate the intrinsic
dipole moment of FQH states due to variations in
 occupation numbers near the edge.
 The boundary is along the direction $\hat {\bm x}$,
and there exists the intrinsic dipole moment $p^y$ proportional to $L$.
We define a function $p^y(k)$ which is the intrinsic dipole moment integrated from the boundary $y = 0$ to $y = k\ell_B^2$,
\begin{equation*}
\frac{p^y(k)}{L} = -e \int_0^k \frac{dk'}{2\pi} \, k' \ell_B^2 (n(k') - \nu).
\end{equation*}
For finite $L$, the integration is approximated by the sum,
\begin{align}
\frac{p^y(k)}{L} &= - \frac{2\pi \ell_B^2 e}{L^2} \times \nonumber \\
&\left(\sum_{m' = 0\text{ or } 1/2}^{m-1}m' \langle n_{m'}\rangle_0+ \textstyle{\frac{1}{2}} m\langle n_m\rangle_0  -\textstyle{\frac{1}{2}}\nu m^2\right)
\end{align}
where the summation is over integers if the state is bosonic or half-integers if fermionic. The last term in the bracket subtracts the
contribution from the uniform density. Because a quantum Hall fluid is
uniform within its bulk, we expect the dipole moment to converge to a value as
$k$ gets large. We also multiply
$p^y(k)/L$ by $(-e/4\pi)^{-1}$ so that it becomes a dimensionless quantity which is
predicted to be $-s/q$, the guiding-center spin divided the number of flux quanta
attached to each composite boson. See Fig.\ref{l2_dipole}, \ref{l3_dipole},
\ref{l4_dipole}, \ref{mr1_dipole} and \ref{mr2_dipole}.
We observe that all intrinsic dipole moments approach expected values
as we integrate up to the center of the fluids.
The expected values are listed in Table.\ref{expected}. Thus, we confirm the relationship (\ref{gc_spin_and_dipole}) between the guiding-center spin and the intrinsic dipole moment holds not only for a droplet but also for the straight edge.


%% file: oes.tex
\subsection{Dipole moment of an ideal edge from orbital entanglement spectrum}

\label{sec:es_dipole}
 We now
describe another method to calculate the guiding center dipole moment using
entanglement spectrum \cite{Li2008}. This gives a connection between the
chirality of entanglement spectrum and the dipole moment.
We denote the full Fock space by $\mc H$, and represent it as
a tensor product of two Fock spaces $\mc H_L$ and $\mc H_R$
so that $\mc H = \mc H_L \otimes \mc H_R$. Given a FQH
ground state $|\Psi\rangle $ in the lowest Landau level, in order to
find an orbital entanglement spectrum,  it can be Schmidt decomposed
\begin{equation}
|\Psi\rangle = \sum_{r} e^{-(1/2)\xi_r} |\Psi_L^r\rangle \otimes |\Psi_R^r\rangle,
\end{equation}
where $|\Psi^r_R\rangle \in \mc H_L$ and $|\Psi^r_R\rangle \in \mc H_R$.
The set of real numbers $\xi_r$ are called the ``entanglement spectrum".
Equivalently, the spectrum can be obtained by diagonalizing the density
matrix of a subsystem
\begin{equation}
\rho_L = \frac{\sum_{r} e^{-\xi_r} |\Psi_L^r\rangle\langle \Psi_L^r|}{\sum_r e^{-\xi_r}}.
\end{equation}
\indent We place a FQH state on a
cylinder, assigning guiding-center momenta. We then divide the whole
system into two subsystems $L$ and $R$ depending on whether
guiding-center momentum quantum number is either positive or negative.
This is known as the ``orbital cut".
The location of  the cut (the zero momentum) does not matter in
principle as long as it belongs to a ``vacuum sector" if the
number of particles is sufficiently large (We clarify what we mean
by vacuum sector later in this section). However, in order to minimize
the finite size effect, we should choose the cut to be located
near the middle of the fluid as much as possible.

For instance, in the case of Laughlin 1/3 state with $N=8$ particles,
we can divide the total system into subsystems as follows,
\begin{equation}
010010010010|010010010010 \nonumber
\end{equation}
In this root occupation, there are $N_L^0 = N_R^0 = 4$ particles in each system.
Assigning zero to the cut, the total guiding-center quantum number
is $M_L^0 = -(\frac{3}{2}+\frac{9}{2}+ \frac{15}{2}+ \frac{21}{2})= -24$ \
for the left subsystem while the right subsystem has $M_R^0 = -M_L^0$.
These are called the ``natural'' values of $N_L, \, N_R ,\, M_L$ and $M_R$.
However, within the subsystem $L$, it can contain any non-negative
 number $N_L$ of particles as long as it satisfies $N_L + N_R = N$.
Also, any total guiding-center quantum number $M_L$ is possible as
long as it satisfies $M_L+ M_R = 0$. The entanglement spectrum
obtained from this orbital cut splits into distinct sectors
labeled by $N_L$ and $M_L$.

The chirality of the entanglement spectrum manifests itself when
we note that the model FQH state derives from a Jack polynomials
so that the many-particle state is spanned by the states obtainable
by squeezing operation. Hence, the Laughlin state is a superposition
of states with $M_L \ge M_L^0$ and $M_R \le M_R^0$ :
\begin{equation}
|\Psi\rangle = \sum_{M_L\ge M_L^0, N_L,r'}
e^{-(1/2)\xi_{r',N_L,M_L}} |\psi_L^{r',N_L, M_L}\rangle \otimes |\psi_R^{r',N_R, M_R}\rangle,
\end{equation}

where  $\{r'\}$ are the remaining labels of states. Thus, the change
in total guiding-center quantum number $\Delta M_L = M_L - M_L^0$ is
always non-negative for any pseudo-energies $\xi_{r}$. We also
define $\Delta N_L = N_L- N_L^0$.

Now, we can calculate the expectation value of $\Delta M_L$
\begin{align}
 \langle\Delta M_L\rangle  = \frac{\sum_{r',N_L,M_L} \Delta M_L e^{-\xi_{r',N_L,M_L}}}{\sum_{r',N_L,M_L} e^{-\xi_{r',N_L,M_L}}}.
\end{align}

 \indent We note that the lower
bound of $\Delta M_L$ is determined by $\Delta N_L$. For instance,
consider the following root occupation,
 \begin{equation*}
 010010010010|01...
 \end{equation*}
By squeezing, we can pull the extra particle on the right side
of the cut and obtain a state in the left subsystem with $\Delta N_L = 1$
and $\Delta M_L = 3/2$ because the shown part of the root occupation has the total
guiding-center quantum number $M_L^0+ 3/2$ and the squeezing
operation does not change the total guiding-center quantum number.
Now, consider the root occupations
 \begin{align*}
  &010010010010|01001...\\
  &010010010000|00000...
 \end{align*}
By similar reasoning, we can obtain a state in the left
subsystem with $\Delta N_L = 2$ and  $\Delta M_L = 3/2+9/2$.
When $\Delta N_L = -1$, it corresponds to the absence of an
 electron with the guiding-center quantum number $-3/2$, and
thus $\Delta M_L = 3/2$. For $1/q$ Laughlin ground states,
with these observations, we can express the quantum number $\Delta M_L$
measured with respect to the ``vacuum" cut as
\begin{equation}
\Delta M_L = \frac{q}{2}(\Delta N_L)^2 + \sum_{m = 0}^{\infty} m b_m^{\dagger}b_m,
\end{equation}
where the boson number operator $ b_m^{\dagger}b_m$  in the second term
can take any non-negative integer values, and it  describes additional increments
in $\Delta M_L$ when the squeezing between a particle in the left subsystem  and
another in the right subsystem does not cause any further change of particle numbers
in each subsystem.  This is exactly the free chiral boson Hamiltonian\cite{Wen1990}.

By the same method, we can deduce that for 2/4 Moore-Read ground
state, $\Delta M_L$ takes a specific form
\begin{align}
\Delta M_L& = \frac{2}{2}(\Delta N_L)^2 + \sum_{m = 0}^{\infty} m b_m^{\dagger}b_m
+ \sum_{m = 1/2}^{\infty} m f_m^{\dagger}f_m  \nonumber\\
&(-1)^{\Delta N_L} = (-1)^{\sum_m  f_m^{\dagger}f_m},
\end{align}
where the second term is the chiral boson contribution, and the last
term is the chiral Majorana fermion contribution. The fermion momenta
are half-integers, and the fermion occupation numbers are either 0 or 1.
The second line is a constraint on the total number of Majorana fermions.
For example,  the minimum change of the total quantum number is $\Delta M_L = 1+1/2$ when $\Delta N_L = 1$.

\subsection{Decomposition of $\langle\Delta M_L\rangle$  and $\langle\Delta N_L\rangle$}
\label{sec:decomposition}

\indent We relate $\langle\Delta M_L\rangle$ to the total
guiding-center momentum $P_x$ of the left subsystem
\begin{equation*}
 P_x = \frac{2\pi \hbar}{L} \langle\Delta M_L\rangle.
\end{equation*}
Furthermore, we relate the momentum to the
dipole moment.
\begin{equation}
p^y = \frac{-e\ell_B^2}{\hbar} P_x = \frac{-2\pi \ell_B^2 e}{L}\langle \Delta M_L\rangle.
\end{equation}
\indent We now show the equivalence of this dipole moment
with that we calculated using occupation numbers.
We can re-write the total guiding-center quantum number $M_L$ as
 \begin{equation}
 M_L = \sum_{m \in  L} m \,n_m,
 \end{equation}
 where $m$'s are the guiding-center quantum numbers
that belong to the left subsystem, and $n_m$'s are the electron
occupation number operators. Meanwhile, $M_L^0$ is just a number
that depends on the root occupation numbers $\{n_m^0\}$ of the model FQHE state
\begin{equation}
M_L^0 = \sum_{m \in  L} m \,n_m^0.
\end{equation}
Then, the expectation value $\langle \Delta M_L\rangle$ is written as
\begin{align}
\langle \Delta M_L\rangle& = \text{Tr}_L\left[\Delta M_L \rho_L \right]\nonumber\\
&= \left(\sum_{m\in L} m \langle n_m\rangle_0 -\bar M_L\right)- (M_L^0 - \bar M_L),
\end{align}
where $\bar M_L = -\nu m_F^2/2$ is the total guiding-center
quantum number for the left subsystem with the uniform number density $\nu$.
The first term is the intrinsic dipole moment calculated previously.
We denote the second term as
 \begin{equation}
 h_{\alpha} = M_L^0 - \bar M_L.
 \end{equation}
$h_{\alpha}$ depends on the location of the cut.
Similarly, we define $N_L^0 = \sum_{m \in L} n_m^0$ and $\bar N_L = \nu | m_F|$.
Then, $\langle \Delta N_L \rangle$ can be written as
\begin{align}
\langle \Delta N_L\rangle &= \text{Tr}_L [\Delta N_L \rho_L]\\
&=  \left(\sum_{m\in L}  \langle  n_m\rangle_0 -\bar N_L\right)- (N_L^0 - \bar N_L).
\end{align}
The first term vanishes by Luttinger's theorem.

We denote the second term as
 \begin{equation}
 q_{\alpha} = N_L^0 - \bar N_L.
 \end{equation}
\indent We can calculate $h_{\alpha}$ and $q_{\alpha}$ for different model FQH
states only using the root occupation numbers. For 1/3 Laughlin state, consider
following different locations of cuts corresponding to quasi-particle,
vacuum and quasi-hole sectors respectively,
\begin{align*}
 &01001001001|001001...&\quad& h_{p} = 1/6&\, &q_{p} = 1/3&\\
 &010010010010|01001...&\quad &h_{\mathbb I} = 0&\, &q_{\mathbb I} = 0&\\
 &0100100100100|1001...&\quad &h_{h} = 1/6& &q_{h} =- 1/3&.
\end{align*}
 For 1/5 Laughlin state, consider following different locations of cuts
corresponding to two and one quasi-particles, vacuum, one and two
quasi-hole sectors respectively,
\begin{align*}
 &010000100001|0000100001...&\,& h_{2p} = 2/5&\, &q_{2p} = 2/5&\\
 &0100001000010|000100001...&\, &h_{p} = 1/10&\, &q_{p} = 1/5&\\
 &01000010000100|00100001...&\, &h_{\mathbb I} = 0& &q_{\mathbb I} =0&\\
 &010000100001000|0100001...&\,& h_{h} = 1/10&\, &q_{h} = -1/5&\\
 &0100001000010000|100001...&\, &h_{2h} = 2/5&\, &q_{2h} =- 2/5&.
\end{align*}
For $2/4$ Moore-Read ground state, we have the following possibilities of
cuts corresponding to isolated fermion,  quasi-particle pair, vacuum  and
quasi-hole pair sectors respectively,
\begin{align*}
 &0110011001|100110...&\quad& h_{\psi} = 1/2&\quad &q_{\psi} = 0&\\
 &01100110011|00110...&\quad& h_{2p} = 1/4&\quad &q_{2p} = 1/2&\\
 &011001100110|0110...&\quad& h_{\mathbb I} = 0&\quad &q_{\mathbb I} = 0&\\
 &0110011001100|110...&\quad& h_{2h} = 1/4&\quad &q_{2h} = -1/2&.
\end{align*}
For $2/4$ Moore-Read state with a quasi-hole at the left Fermi  surface,
\begin{align*}
 &010101010101|0101...&\quad& h_{p} = 1/16&\quad &q_{p} = 1/4&\\
 &0101010101010|101...&\quad& h_{h} = 1/16&\quad &q_{h} = -1/4&.
\end{align*}
We see that $h_{\alpha}$ are exactly the conformal spins of the
elementary excitations ($h_{\alpha}$ were called ``topological spin"
by other authors\cite{Qi2012,Zaletel2012}. $q_{\alpha}$ are the fractional
charge of the elementary excitations.

\begin{figure}[htb]
\includegraphics[width=0.85\linewidth, page = 24]{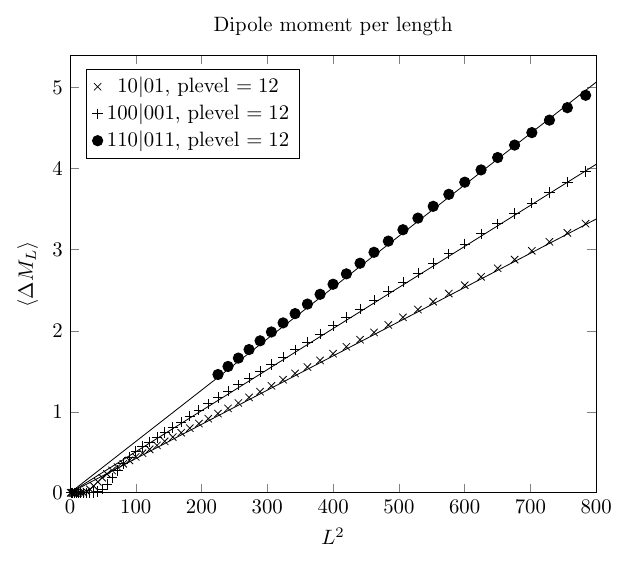}
\caption{ Dots represent $\langle \Delta M_L\rangle$ of 1/3 Laughlin
($\times$), 1/5 Laughlin (+) and 2/4 Moore-Read ($\bullet$) states
for different values of circumference $L$ calculated from the orbital
entanglement spectra with the truncation level equal to 12. Here, each
orbital cut is a vacuum cut, $h_{\alpha} =0$. Each line represents the
dominant value $L^2/(8\pi^2\ell_B^2)\times (-s/q)$ where $-s/q = 1/3, \,2/5,\, 1/2$
respectively.
}
\label{mean_virasoro}
\end{figure}

\begin{figure}[htb]
\includegraphics[width=0.85\linewidth, page = 25]{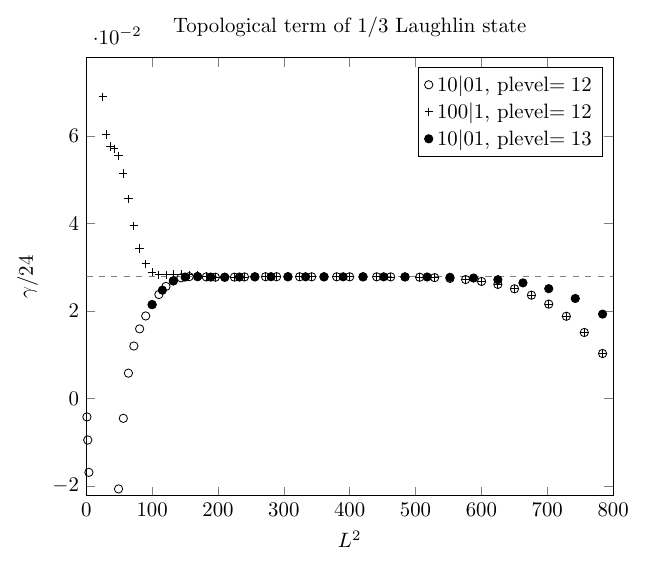}
\caption{The plot of sub-leading term $\gamma/24$ for 1/3 Laughlin state. $\circ$:
vacuum cut with truncation level 12. +:  quasi-hole cut with truncation level 12. $\bullet$:
vacuum cut with truncation level 13. The horizontal line represents 1/36.}
\label{lau3_topo}
\end{figure}

\begin{figure}[htb]
\includegraphics[width=0.85\linewidth, page = 26]{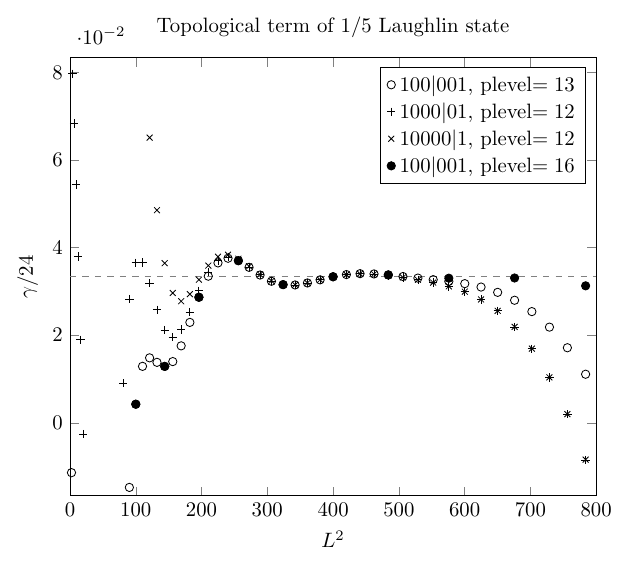}
\caption{The plot of sub-leading term $\gamma/24$ for 1/5 Laughlin state. $\circ$:
 vacuum cut with truncation level 13. +: one quasi-hole cut with truncation level 12.
$\times$: two quasi-hole cut with truncation level 12.  $\bullet$: vacuum cut with truncation level 16.
The horizontal line represents 1/30.}
\label{lau5_topo}
\end{figure}

\begin{figure}[htb]
\includegraphics[width=0.85\linewidth, page = 27]{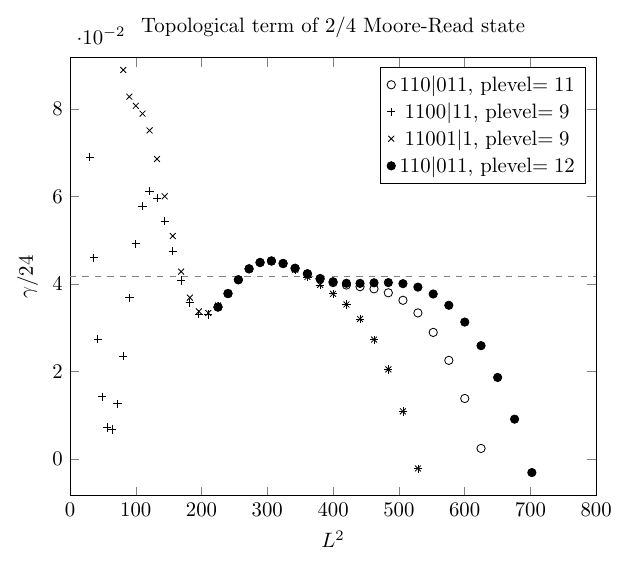}
\caption{The plot of sub-leading term $\gamma/24$ for 2/4 Moore-Read state.
$\circ$:  vacuum cut with truncation level 11. +: one quasi-hole cut with
truncation level 9. $\times$: isolated fermion cut with truncation level 9.
$\bullet$: vacuum cut with truncation level 12. The horizontal line represents 1/24.}
\label{mr2_topo}
\end{figure}

In $\langle \Delta M_L\rangle$, the most dominant term is proportional
to the squared circumference $L^2$. We define the sub-leading term as
\begin{equation}
\frac{\gamma}{24} - h_{\alpha}= \langle \Delta M_L\rangle + \frac{1}{2}\left(\frac{L}{2\pi \ell_B}\right)^2 \frac{s}{q}
\end{equation}
\indent In order to calculate this sub-leading term, we need a large system.
We generated orbital entanglement spectra for 1/3 Laughlin state, 1/5 Laughlin state and 2/4
Moore-Read state using the ``matrix product state'' program developed by
Regnault \textit{et al}\cite{Regnault2013}. Each state contains 100 particles.
Their accuracy is limited by the so-called  ``truncation level'' (which we
call plevel in the figures). As the truncation level increases the approximation
to the exact state gets better. We plot the $\langle \Delta M_L\rangle$ against
different values of circumference in Fig.\ref{mean_virasoro}.
The sub-leading term $\gamma/24$ is also plotted in
 Fig.\ref{lau3_topo}, \ref{lau5_topo} and \ref{mr2_topo}.
The numerical calculation is consistent with the prediction\cite{newpaper}
 that $\gamma$ may be expressed as
\begin{equation}
  \gamma =  \tilde c -  \nu
\label{fundamental} \end{equation}
where $\tilde c$ is the total \textit{signed} central
charge  $c-\bar{c}$  of the underlying edge theory:
$\tilde c=1$ for Laughlin states and $\tilde c= 3/2$ for 2/4 Moore-Read  state.

The theoretical derivation of this result will be presented elsewhere\cite{newpaper}.
It is the anomaly of the \textit{signed}  Virasoro algebra\cite{newpaper},
with generators $\tilde L_m$ = $L_m - \bar L_{-m}$, which  are the Fourier components
of the momentum density; this
survives as a universal algebra, with no renormalization,  despite the breaking of Lorentz
and conformal invariance when the various linearly-dispersing modes acquire different
propagation speeds.   Note that integer quantum Hall states, where the effect is due to
simple filling of Landau levels by the Pauli principle (and which are \textit{not}
topologically ordered) do not exhibit a gapless ``orbital'' entanglement spectrum of
the type
discussed here, and have $\tilde c - \nu$ = 0.    The anomaly $\tilde c$
appears in (\ref{fundamental})
as a ``Casimir momentum'' , which is a feature of chiral
theories: this remains universal so long as translational invariance is
unbroken, while the
Casimir energy (the origin of the  finite-size correction in
non-chiral cft) becomes non-universal once Lorentz invariance is lost.


%% file: res.tex
\subsection{Momentum polarization from the real-space cut}
\label{sec:ent_realsp}
The ``orbital-cut'' entanglement spectrum only has a gapless spectrum
when it is applied to states with topological order.   In particular, it does
\textit{not} show a gapless spectrum when applied to integer quantum Hall
states, which are not topologically-ordered (they do not exhibit a topological
 ground-state degeneracy when constructed on surfaces on genus $> 0$,
which is the defining property of ``topological order'').
Dubail \textit{et al.}\cite{Dubail2012} perceived this feature as a defect of the
orbital-cut method, and
introduced a modified ``real-space'' entanglement spectrum  for quantum
Hall states as a  remedy.
 (However, it should be noted that the absence of a gapless orbital-cut
entanglement spectrum in the trivial integer QHE case
is consistent with Li and Haldane's
claim\cite{Li2008} that a gapless spectrum is a characteristic
property of a topologically-ordered state.)

In the high field limit, quantum Hall states in Landau levels become an
unentangled product of the state of the guiding-centers $\bm R_i$
and the Landau orbit (cyclotron motion) radii $\tilde {\bm R_i}$.
 Each Landau level is
characterized by a form-factor
\begin{equation}
f_n(\bm q) = \langle \psi_n | e^{i\bm q\cdot \tilde {\bm R}}| \psi_n \rangle_L
= 1 - \tilde \Lambda_n^{ab}q_aq_b\ell_B^2 + O(q^4),
\end{equation}
where $|\psi_n\rangle_L$ is the $n$-th Landau level single-particle state.
If only a single Landau level is occupied, the electronic state is
a simple product of the guiding center state used in the ``orbital cut''
with a trivial completely-symmetric state of the Landau-orbit radii,
characterized by a form factor $f(\bm q)$ = $f(q_x,q_y)$.
(This is the type of state for which
the ``real-space cut'' was constructed in \cite{Dubail2012}.)
In the ``Landau gauge'', the wavefunctions $\phi_m(x,y)$ have a profile
\begin{equation}
|\phi_{n,m}(x,y)|^2 = \frac{1}{ L}\int_{-\infty}^{\infty} \frac{dq_y}{2\pi} f_n(0,q_y)e^{iq_y(y-y_m)} ,
\end{equation}
where  $y_m$ =  $2\pi m \ell_B^2/L$, $m \in \mathbb Z+1/2$.
The real-space cut at $y = 0$
is based on the partition
\begin{equation}
P^L_{n,m} = \int _{-\infty}^{0} dy\int_0^Ldx |\phi_{n,m}(x,y)|^2, \quad P^L_{n,m}+P^R_{n,m} = 1.
\label{rsp}
\end{equation}
 Note also that
\begin{align}
 &\sum_{m> 0}   m P_{n,m}^L -\sum_{m<0} m P^R_{n,m} = \sum_m m (P_{n,m}^L -\theta(m)) \nonumber\\
 &\quad=  \frac{\tilde \Lambda_n^{yy}L^2}{(2\pi\ell_B)^2} + \frac{1}{24} + \mc O(L^{-1}). \label{real_sp_dipole}
\end{align}
where for Galilean-invariant Landau levels with an effective mass
tensor $m\tilde g_{ab}$ (with $\det \tilde g$ = 1),
\begin{equation}
\tilde \Lambda_n^{ab} =  \frac{1}{2}\tilde s_n\tilde g^{ab}.
\end{equation}

In order to obtain the dipole moment from the real-space
cut\cite{Dubail2012}, we first double
the single-particle Hilbert space $\mathcal H_1$ on a cylinder
into two subspaces $\mathcal H_{1L}$ and $\mathcal H_{1R}$
where a new ``pseudospin'' index that takes values ``$R$'' and
 ``$L$'' has been introduced:
\begin{equation}
\mathcal H_1 \mapsto \mathcal H_{1L} \otimes \mathcal H_{1R}.
\end{equation}
If a function $f(\bm r)$ belongs
to $\mc H_{1X}$, then $f(\bm r) = 0 $ if $\bm r \not\in X$
where $X$ can be either the subsystem $L$ or $R$.
We choose the line $x = 0$ to be the boundary along the
translational invariant direction so that the guiding-center
remains as a good quantum number.  Now, consider the Fock
space $\mc H$. Denote a vacuum state with no particle by $|\text{vac}\rangle$.
We create a particle with the guiding-center $m$ in $n$-th Landau level
by $c_{n,m}^{\dagger}$. This creation operator can be decomposed as
\begin{subequations}
 \begin{align}
c_{n,m}^{\dagger}& =  u_{n,m} c^{\dagger}_{n,m,L} + v_{n,m} c^{\dagger}_{n,m,R}\\
|u_{n,m}|^2  &= P_{n,m}^L, \quad  |v_{n,m}|^2 = P^R_{n,m} ,
\end{align}
\end{subequations}
where the physical state satisfies the constraint
\begin{equation}
(v_{n,m}c_{n,m,L} - u_{n,m} c_{n,m,R})|\Psi \rangle = 0,
\end{equation}
so all occupied orbitals have a  pseudospin which is
fully-polarized in the ``physical'' direction.
For notational convenience, we concentrate on a single
Landau level and drop the index $n$. Given a Slater determinant
state $|\{n_m\}\rangle$ labeled by occupation numbers $n_m$,
\begin{align}
&|\{n_m\}\rangle = \prod_m (c_m^{\dagger})^{n_m} |\text{vac}\rangle \nonumber\\
&=  \prod_m \left(u_mc^{\dagger}_{m ,L} +  v_mc^{\dagger}_{m, R}\right)^{n_m} |\text{vac}\rangle
\end{align}
the product of creation operators can be expanded.
Then, we obtain
\begin{equation}
|\{n_m\}\rangle = \sum_{\alpha,\beta :N_L+N_R = N}
A_{\alpha\beta}(\{n_m\})|\Psi_{\alpha}^L\rangle \otimes |\Psi_{\beta}^R\rangle
\end{equation}
where $|\Psi_{\alpha}^X\rangle$ are Slater determinant states
belonging to the Fock space $\mc H_X$ ($X = L,\, R$) and $A_{\alpha\beta}(\{n_m\})$
is a product of $u_m$
and $v_m$ . With this expansion, and after translating
the partition $\lambda$ into the occupation numbers $\{n_m\}$,
the mapping of a Jack polynomial into a model FQH
state $|\Psi\rangle$ in (\ref{jack}) becomes
\begin{equation}
|\Psi\rangle = \sum_{\stackrel{\alpha,\beta}{N_L+N_R = N}} \sum_{\{n_m\} \le \{n_m^0\}}
a_{\{n_m\}} A_{\alpha\beta}(\{n_m\})|\Psi_{\alpha}^L\rangle \otimes |\Psi_{\beta}^R\rangle
\end{equation}
\indent We can further Schmidt-decompose the model FQH state $|\Psi\rangle$.
However, if our objective is only to calculate the diagonal operators such as $ M_L $
and $N_L$, the information we gathered from the orbital cut is enough.
 Consider the expectation value of the operator $n_m^L = c^{\dagger}_{m,L} c_{m,L}$
\begin{equation}
\langle n_{m}^L\rangle' = \text{Tr}_L[  c_{m,L}^{\dagger} c_{m,L} \rho_L']
\end{equation}
where $\rho_L'$ is the normalized density matrix for the subsystem
$L$, and we placed an apostrophe on the bracket $\langle ...\rangle'$
to distinguish the real-space cut expectation value with the orbital cut
expectation value $\langle...\rangle$. For all guiding-centers $m'$
such that $m' \neq m$, the factors $P_{m'}^L$ and $P_{m'}^R$ appear in
pairs in the expectation value,  and add to one. From this observation,
we see that the expectation value simplifies to
\begin{equation}
\langle  n_m^L \rangle'  = P_m^L\langle n_m\rangle_0
\end{equation}
\indent Using this expression, in the expectation
value of $\Delta M_L $,
\begin{align}
\langle \Delta M_L\rangle' &= \sum_m m \langle n^L_m \rangle' - M_L^0 \nonumber \\
&= \sum_m m (P_m^L -\theta(m))\langle n_m \rangle_0  \nonumber\\
&\quad+ \sum_{m <0} m \langle n_m \rangle_0- M_L^0.
\end{align}
The first term is an additional term that appears when we
consider the real-space cut. The second term is the expectation
value of $\Delta M_L$ with the orbital cut that we calculated previously.
In the first term, $P_m^L$ $\rightarrow 1$ for $m \ll 0$, and the summand vanishes.
Meanwhile, as $m \rightarrow 0$, which is the location of the real-space cut,
we are deep into the bulk so that $\langle n_m\rangle_0 = \nu$.
Thus, in the thermodynamic limit, the expectation value $\langle \Delta M_L\rangle'$ becomes
\begin{align}
\langle \Delta M_L\rangle' &= \nu \sum_m m (P_m^L -\theta(m))
+ \langle \Delta M_L\rangle
\end{align}
 The first term was already considered in (\ref{real_sp_dipole}). \\

\indent For simplicity, we now assume Galilean-invariant Landau orbits, so
$\tilde \Lambda_n^{yy}$ = ${\textstyle\frac{1}{2}}\tilde s_n\tilde g^{yy}$,
where $\tilde s_n$  = $n +{\textstyle\frac{1}{2}}$ is the Landau-orbit
spin. If we further include the contributions from the filled Landau levels
0,1,...,$n-1$, then $\langle \Delta M_L\rangle'$ is
\begin{align}
\langle \Delta M_L\rangle' =&
\frac{1}{2}\left(\frac{L}{2\pi \ell_B}\right)^2  \left(\sum_{n'=0}^n
\tilde s_{n'}\nu_{n'}\tilde g^{yy}- \frac{s}{q}g^{yy}\right)\nonumber \\
+& \frac{(\nu' + \gamma) }{24} + \mc O(L^{-1}),
\label{mainresult}
\end{align}
where $\nu_{n'}$ = $\tilde c_{n'}$ =  1 for $n' < n$ and $\nu_n = \nu$. We also defined $\nu' = \sum_{n'=0}^n\nu_{n'}$ and $\gamma$ = $\tilde c_n - \nu_n$.
We explicitly wrote the two metrics $g_{ab}$ and $\tilde g_{ab}$
since they need not coincide as noted before \cite{Haldane2011}.
There are topological contributions from each cut: we get $n/24$
from $n$ filled Landau levels and  $\nu/24$ from the partially filled Landau level as a result of the real-space cut. We get $\gamma/24$ from the variation of orbital occupations near the physical edge. The normal vector of the surface of the fluid at the physical edge is reversed from the normal vector at the real-space cut. We note here that the Landau-orbit spins $\tilde s_{n'}$ ($n' = 0,\dots, n$) are positive while the guiding-center spin $s$ is negative. The general expression for the total Hall viscosity tensor $\eta_H'^{ab}$ (the sum of the Landau-orbit ant guiding-center contributions) is
\begin{subequations}
\begin{align}
\eta_H'^{abcd} =&
{\textstyle\frac{1}{2}}\left (\eta_H'^{ac}\epsilon^{bd} + \eta_H'^{bd}\epsilon^{ac}
+ \eta_H'^{bc}\epsilon^{ad} + \eta_H'^{ad}\epsilon^{bc}\right)\\
\eta_H'^{ab} =& \frac{eB}{2\pi}\left ( \sum_n \tilde \Lambda_n^{ab}\nu_n -\frac{1}{2}\frac{s}{q} g^{ab}\right ) ,
\end{align}
\label{Hall_viscosity}
\end{subequations}
 Using this expression for the Hall viscosity, we can write the momentum polarization in a fully covariant tensor form as
\begin{equation}
 \langle \Delta M_L\rangle'
=  \hbar^{-1} \eta'^{ab}_H\epsilon_{ac}\epsilon_{bd}\frac{L^cL^d}{2\pi\ell_B^2} +  \frac{(\nu' + \gamma) }{24}
 \end{equation}

The  $O(L^2)$ term gives the Hall viscosity, which  is now the sum of two terms:
one is derived from the Landau-orbit form factors, weighted by the
Landau level occupation, and the other is the guiding-center contribution derived from the orbital cut.

We note the the ``real-space cut'' involves far greater  computational effort
than the ``orbital cut'', but at least as far as the  ``momentum polarization'' is concerned,
merely adds trivial contributions to the Hall viscosity and topological terms
\textit{e.g.},  $(\tilde c - \nu) + \nu = \tilde c$.    Clearly all the non-trivial topological
and entanglement  information  of the  topologically-ordered states is fully
present in the ``orbital-cut''.   From this viewpoint, we are tempted to conclude
that use of the ``real-space cut''  is an unnecessary  use
of computational resources that merely serves to conceal the structures of
the ``orbital cut'' entanglement spectrum by convoluting them with  the form-factor of the Landau orbits.

%% file: conclusion.tex
\section{Conclusion}
We showed that the intrinsic dipole moment  along the edges of the
incompressible FQH fluids can be expressed in terms of electric charge $e$,
guiding center spin $s$, number of fluxes per a composite boson $q$,
confirming the prediction made in the previous work \cite{Haldane2009}.
This provides another sum rule for the FQH fluids in addition to the
Luttinger sum rule\cite{Luttinger1960}. For incompressible FQH states, the electric force on the intrinsic dipole moment is balanced the stress
given by the gradient of the flow velocity times the guiding-center Hall viscosity.

We also related the
the edge dipole moment to the  expectation value of the momentum
(or ``momentum polarization''\cite{Qi2012})
of the entanglement spectrum.      In the high-field limit, when the
guiding-center and Landau-orbit degrees of freedom become
unentangled with each other,
the  dipole moment and the related Hall viscosity separate cleanly
into independent parts respectively coming from the non-trivial
correlated guiding-center degrees of
freedom of the FQH state, and the trivially-calculable one-body
properties of the Landau orbits.      The ``orbital cut'' entanglement
spectrum introduced by Li and Haldane\cite{Li2008} contains only
information on the guiding-center degrees of freedom, and allows the
guiding-center contribution to the Hall viscosity of the FQH fluid to be
found as a bulk geometric property, and also  gives the topological
quantity $\gamma$ = $\tilde c - \nu$, the difference between the   (signed) ``conformal anomaly''
(or ``chiral stress-energy anomaly''\cite{newpaper} $\tilde c$ = $c-\bar c$,
and the chiral charge anomaly $\nu$, which are the two fundamental quantum anomalies
of the FQH fluids.      It is useful to note that $\gamma$ is
insensitive to completely-filled Landau levels, and vanishes identically in
integer quantum Hall states, which do not exhibit topological-order.

We also examined the equivalent calculation in the ``real-space''
entanglement spectrum described by Dubail \textit{et al.}\cite{Dubail2012}, which adds information
about the Landau orbit to  provide the combined guiding-center plus
Landau-orbit contribution to the Hall viscosity and $\tilde c$ rather than $\gamma$.
However since the  ``real-space entanglement''
method involves much extra computational complexity, and convolutes
the non-trivial Landau-orbit-independent correlated guiding center data
with the essentially trivial
(and Landau-level-dependent)   Landau-orbit form factor data,  we concluded
that there were no advantages to use of  the ``real-space'' as opposed to ``orbital''
entanglement spectrum.    Indeed,  since the Landau-orbit  form factor is essentially
unrelated to the FQH correlations,  and can be chosen as an additional
(and arbitrary) ingredient  to convert orbital entanglement data into a ``real-space''
form, its use may actually serve to conceal the essential features of the guiding-center entanglement.
The ``real-space'' spectrum may also be thought of operationally as the use of an
essentially \textit{ad-hoc} function $P^L_m$  (\ref{rsp}) that can be arbitarily
chosen to ``smear out'' a sharp orbital cut between cylinder orbitals $m$ and $m+1$,
which breaks both guiding-center indistiguishability (by introducing ``pseudo-spin''
labels ``$L$'' and ``$R$'')
and reducing the full 2D  translational symmetry (the parallel to the cylinder axis
(in the $N\rightarrow \infty$ limit, or  equivalently, full rotational symmetry in the
spherical geometry) to 1D axial translational symmetry.
It interpolates continuously between
two completely-well-defined limits of guiding-center entanglement:
the ``orbital cut'' which preserves guiding-center indistinguishability
while breaking 2D translational symmetry down to 1D translational
symmetry, and the ``particle cut'' which divides the guiding centers into
two distinguishable groups, but preserves full 2D translational symmetry.

Acknowledgements: This work was supported in part  by the Department of Energy,
Office of Basic Energy Sciences through grant No.
DE-SC0002140 and also by the W. M. Keck foundation.

%% file: edge_dipole.bbl
\begin{thebibliography}{28}%
\makeatletter
\providecommand \@ifxundefined [1]{%
 \@ifx{#1\undefined}
}%
\providecommand \@ifnum [1]{%
 \ifnum #1\expandafter \@firstoftwo
 \else \expandafter \@secondoftwo
 \fi
}%
\providecommand \@ifx [1]{%
 \ifx #1\expandafter \@firstoftwo
 \else \expandafter \@secondoftwo
 \fi
}%
\providecommand \natexlab [1]{#1}%
\providecommand \enquote  [1]{``#1''}%
\providecommand \bibnamefont  [1]{#1}%
\providecommand \bibfnamefont [1]{#1}%
\providecommand \citenamefont [1]{#1}%
\providecommand \href@noop [0]{\@secondoftwo}%
\providecommand \href [0]{\begingroup \@sanitize@url \@href}%
\providecommand \@href[1]{\@@startlink{#1}\@@href}%
\providecommand \@@href[1]{\endgroup#1\@@endlink}%
\providecommand \@sanitize@url [0]{\catcode `\\12\catcode `\$12\catcode
  `\&12\catcode `\#12\catcode `\^12\catcode `\_12\catcode `\%12\relax}%
\providecommand \@@startlink[1]{}%
\providecommand \@@endlink[0]{}%
\providecommand \url  [0]{\begingroup\@sanitize@url \@url }%
\providecommand \@url [1]{\endgroup\@href {#1}{\urlprefix }}%
\providecommand \urlprefix  [0]{URL }%
\providecommand \Eprint [0]{\href }%
\providecommand \doibase [0]{http://dx.doi.org/}%
\providecommand \selectlanguage [0]{\@gobble}%
\providecommand \bibinfo  [0]{\@secondoftwo}%
\providecommand \bibfield  [0]{\@secondoftwo}%
\providecommand \translation [1]{[#1]}%
\providecommand \BibitemOpen [0]{}%
\providecommand \bibitemStop [0]{}%
\providecommand \bibitemNoStop [0]{.\EOS\space}%
\providecommand \EOS [0]{\spacefactor3000\relax}%
\providecommand \BibitemShut  [1]{\csname bibitem#1\endcsname}%
\let\auto@bib@innerbib\@empty
\bibitem [{\citenamefont {Tsui}\ \emph {et~al.}(1982)\citenamefont {Tsui},
  \citenamefont {Stormer},\ and\ \citenamefont {Gossard}}]{Tsui1982}%
  \BibitemOpen
  \bibfield  {author} {\bibinfo {author} {\bibfnamefont {D.~C.}\ \bibnamefont
  {Tsui}}, \bibinfo {author} {\bibfnamefont {H.~L.}\ \bibnamefont {Stormer}}, \
  and\ \bibinfo {author} {\bibfnamefont {A.~C.}\ \bibnamefont {Gossard}},\
  }\href {\doibase 10.1103/PhysRevLett.48.1559} {\bibfield  {journal} {\bibinfo
   {journal} {Phys. Rev. Lett.}\ }\textbf {\bibinfo {volume} {48}},\ \bibinfo
  {pages} {1559} (\bibinfo {year} {1982})}\BibitemShut {NoStop}%
\bibitem [{\citenamefont {Avron}\ \emph {et~al.}(1995)\citenamefont {Avron},
  \citenamefont {Seiler},\ and\ \citenamefont {Zograf}}]{Avron1995}%
  \BibitemOpen
  \bibfield  {author} {\bibinfo {author} {\bibfnamefont {J.~E.}\ \bibnamefont
  {Avron}}, \bibinfo {author} {\bibfnamefont {R.}~\bibnamefont {Seiler}}, \
  and\ \bibinfo {author} {\bibfnamefont {P.~G.}\ \bibnamefont {Zograf}},\
  }\href {\doibase 10.1103/PhysRevLett.75.697} {\bibfield  {journal} {\bibinfo
  {journal} {Phys. Rev. Lett.}\ }\textbf {\bibinfo {volume} {75}},\ \bibinfo
  {pages} {697} (\bibinfo {year} {1995})}\BibitemShut {NoStop}%
\bibitem [{\citenamefont {Read}(2009)}]{Read2009}%
  \BibitemOpen
  \bibfield  {author} {\bibinfo {author} {\bibfnamefont {N.}~\bibnamefont
  {Read}},\ }\href {\doibase 10.1103/PhysRevB.79.045308} {\bibfield  {journal}
  {\bibinfo  {journal} {Phys. Rev. B}\ }\textbf {\bibinfo {volume} {79}},\
  \bibinfo {pages} {045308} (\bibinfo {year} {2009})}\BibitemShut {NoStop}%
\bibitem [{\citenamefont {Read}\ and\ \citenamefont {Rezayi}(2011)}]{Read2011}%
  \BibitemOpen
  \bibfield  {author} {\bibinfo {author} {\bibfnamefont {N.}~\bibnamefont
  {Read}}\ and\ \bibinfo {author} {\bibfnamefont {E.~H.}\ \bibnamefont
  {Rezayi}},\ }\href {\doibase 10.1103/PhysRevB.84.085316} {\bibfield
  {journal} {\bibinfo  {journal} {Phys. Rev. B}\ }\textbf {\bibinfo {volume}
  {84}},\ \bibinfo {pages} {085316} (\bibinfo {year} {2011})}\BibitemShut
  {NoStop}%
\bibitem [{\citenamefont {Haldane}(2009)}]{Haldane2009}%
  \BibitemOpen
  \bibfield  {author} {\bibinfo {author} {\bibfnamefont {F.~D.~M.}\
  \bibnamefont {Haldane}},\ }\href@noop {} {\  (\bibinfo {year} {2009})},\
  \Eprint {http://arxiv.org/abs/0906.1854} {arXiv:0906.1854} \BibitemShut
  {NoStop}%
\bibitem [{\citenamefont {Haldane}(2011)}]{Haldane2011}%
  \BibitemOpen
  \bibfield  {author} {\bibinfo {author} {\bibfnamefont {F.~D.~M.}\
  \bibnamefont {Haldane}},\ }\href@noop {} {\bibfield  {journal} {\bibinfo
  {journal} {Phys. Rev. Lett.}\ }\textbf {\bibinfo {volume} {107}},\ \bibinfo
  {pages} {116801} (\bibinfo {year} {2011})}\BibitemShut {NoStop}%
\bibitem [{\citenamefont {Hoyos}\ and\ \citenamefont {Son}(2012)}]{Son2012}%
  \BibitemOpen
  \bibfield  {author} {\bibinfo {author} {\bibfnamefont {C.}~\bibnamefont
  {Hoyos}}\ and\ \bibinfo {author} {\bibfnamefont {D.~T.}\ \bibnamefont
  {Son}},\ }\href {\doibase 10.1103/PhysRevLett.108.066805} {\bibfield
  {journal} {\bibinfo  {journal} {Phys. Rev. Lett.}\ }\textbf {\bibinfo
  {volume} {108}},\ \bibinfo {pages} {066805} (\bibinfo {year}
  {2012})}\BibitemShut {NoStop}%
\bibitem [{\citenamefont {Maciejko}\ \emph {et~al.}(2013)\citenamefont
  {Maciejko}, \citenamefont {Hsu}, \citenamefont {Kivelson}, \citenamefont
  {Park},\ and\ \citenamefont {Sondhi}}]{Maciejko2013}%
  \BibitemOpen
  \bibfield  {author} {\bibinfo {author} {\bibfnamefont {J.}~\bibnamefont
  {Maciejko}}, \bibinfo {author} {\bibfnamefont {B.}~\bibnamefont {Hsu}},
  \bibinfo {author} {\bibfnamefont {S.~A.}\ \bibnamefont {Kivelson}}, \bibinfo
  {author} {\bibfnamefont {Y.}~\bibnamefont {Park}}, \ and\ \bibinfo {author}
  {\bibfnamefont {S.~L.}\ \bibnamefont {Sondhi}},\ }\href {\doibase
  10.1103/PhysRevB.88.125137} {\bibfield  {journal} {\bibinfo  {journal} {Phys.
  Rev. B}\ }\textbf {\bibinfo {volume} {88}},\ \bibinfo {pages} {125137}
  (\bibinfo {year} {2013})}\BibitemShut {NoStop}%
\bibitem [{\citenamefont {Bradlyn}\ \emph {et~al.}(2012)\citenamefont
  {Bradlyn}, \citenamefont {Goldstein},\ and\ \citenamefont
  {Read}}]{Bradlyn2012}%
  \BibitemOpen
  \bibfield  {author} {\bibinfo {author} {\bibfnamefont {B.}~\bibnamefont
  {Bradlyn}}, \bibinfo {author} {\bibfnamefont {M.}~\bibnamefont {Goldstein}},
  \ and\ \bibinfo {author} {\bibfnamefont {N.}~\bibnamefont {Read}},\ }\href
  {\doibase 10.1103/PhysRevB.86.245309} {\bibfield  {journal} {\bibinfo
  {journal} {Phys. Rev. B}\ }\textbf {\bibinfo {volume} {86}},\ \bibinfo
  {pages} {245309} (\bibinfo {year} {2012})}\BibitemShut {NoStop}%
\bibitem [{\citenamefont {Li}\ and\ \citenamefont {Haldane}(2008)}]{Li2008}%
  \BibitemOpen
  \bibfield  {author} {\bibinfo {author} {\bibfnamefont {H.}~\bibnamefont
  {Li}}\ and\ \bibinfo {author} {\bibfnamefont {F.~D.~M.}\ \bibnamefont
  {Haldane}},\ }\href {\doibase 10.1103/PhysRevLett.101.010504} {\bibfield
  {journal} {\bibinfo  {journal} {Phys. Rev. Lett.}\ }\textbf {\bibinfo
  {volume} {101}},\ \bibinfo {pages} {010504} (\bibinfo {year}
  {2008})}\BibitemShut {NoStop}%
\bibitem [{\citenamefont {Zaletel}\ \emph {et~al.}(2013)\citenamefont
  {Zaletel}, \citenamefont {Mong},\ and\ \citenamefont
  {Pollmann}}]{Zaletel2012}%
  \BibitemOpen
  \bibfield  {author} {\bibinfo {author} {\bibfnamefont {M.~P.}\ \bibnamefont
  {Zaletel}}, \bibinfo {author} {\bibfnamefont {R.~S.~K.}\ \bibnamefont
  {Mong}}, \ and\ \bibinfo {author} {\bibfnamefont {F.}~\bibnamefont
  {Pollmann}},\ }\href {\doibase 10.1103/PhysRevLett.110.236801} {\bibfield
  {journal} {\bibinfo  {journal} {Phys. Rev. Lett.}\ }\textbf {\bibinfo
  {volume} {110}},\ \bibinfo {pages} {236801} (\bibinfo {year}
  {2013})}\BibitemShut {NoStop}%
\bibitem [{\citenamefont {{Tu}}\ \emph {et~al.}(2013)\citenamefont {{Tu}},
  \citenamefont {{Zhang}},\ and\ \citenamefont {{Qi}}}]{Qi2012}%
  \BibitemOpen
  \bibfield  {author} {\bibinfo {author} {\bibfnamefont {H.-H.}\ \bibnamefont
  {{Tu}}}, \bibinfo {author} {\bibfnamefont {Y.}~\bibnamefont {{Zhang}}}, \
  and\ \bibinfo {author} {\bibfnamefont {X.-L.}\ \bibnamefont {{Qi}}},\ }\href
  {\doibase 10.1103/PhysRevB.88.195412} {\bibfield  {journal} {\bibinfo
  {journal} {\prb}\ }\textbf {\bibinfo {volume} {88}},\ \bibinfo {eid} {195412}
  (\bibinfo {year} {2013})}\BibitemShut {NoStop}%
\bibitem [{\citenamefont {Wen}(1991)}]{Wen1990}%
  \BibitemOpen
  \bibfield  {author} {\bibinfo {author} {\bibfnamefont {X.~G.}\ \bibnamefont
  {Wen}},\ }\href {\doibase 10.1103/PhysRevB.43.11025} {\bibfield  {journal}
  {\bibinfo  {journal} {Phys. Rev. B}\ }\textbf {\bibinfo {volume} {43}},\
  \bibinfo {pages} {11025} (\bibinfo {year} {1991})}\BibitemShut {NoStop}%
\bibitem [{\citenamefont {Bernevig}\ and\ \citenamefont
  {Haldane}(2008)}]{Bernevig2008}%
  \BibitemOpen
  \bibfield  {author} {\bibinfo {author} {\bibfnamefont {B.~A.}\ \bibnamefont
  {Bernevig}}\ and\ \bibinfo {author} {\bibfnamefont {F.~D.~M.}\ \bibnamefont
  {Haldane}},\ }\href {\doibase 10.1103/PhysRevLett.100.246802} {\bibfield
  {journal} {\bibinfo  {journal} {Phys. Rev. Lett.}\ }\textbf {\bibinfo
  {volume} {100}},\ \bibinfo {pages} {246802} (\bibinfo {year}
  {2008})}\BibitemShut {NoStop}%
\bibitem [{\citenamefont {Dubail}\ \emph {et~al.}(2012)\citenamefont {Dubail},
  \citenamefont {Read},\ and\ \citenamefont {Rezayi}}]{Dubail2012}%
  \BibitemOpen
  \bibfield  {author} {\bibinfo {author} {\bibfnamefont {J.}~\bibnamefont
  {Dubail}}, \bibinfo {author} {\bibfnamefont {N.}~\bibnamefont {Read}}, \ and\
  \bibinfo {author} {\bibfnamefont {E.~H.}\ \bibnamefont {Rezayi}},\ }\href
  {\doibase 10.1103/PhysRevB.85.115321} {\bibfield  {journal} {\bibinfo
  {journal} {Phys. Rev. B}\ }\textbf {\bibinfo {volume} {85}},\ \bibinfo
  {pages} {115321} (\bibinfo {year} {2012})}\BibitemShut {NoStop}%
\bibitem [{\citenamefont {Laughlin}(1983)}]{Laughlin1983}%
  \BibitemOpen
  \bibfield  {author} {\bibinfo {author} {\bibfnamefont {R.~B.}\ \bibnamefont
  {Laughlin}},\ }\href {\doibase 10.1103/PhysRevLett.50.1395} {\bibfield
  {journal} {\bibinfo  {journal} {Phys. Rev. Lett.}\ }\textbf {\bibinfo
  {volume} {50}},\ \bibinfo {pages} {1395} (\bibinfo {year}
  {1983})}\BibitemShut {NoStop}%
\bibitem [{\citenamefont {Zhang}\ \emph {et~al.}(1989)\citenamefont {Zhang},
  \citenamefont {Hansson},\ and\ \citenamefont {Kivelson}}]{Zhang1989}%
  \BibitemOpen
  \bibfield  {author} {\bibinfo {author} {\bibfnamefont {S.~C.}\ \bibnamefont
  {Zhang}}, \bibinfo {author} {\bibfnamefont {T.~H.}\ \bibnamefont {Hansson}},
  \ and\ \bibinfo {author} {\bibfnamefont {S.}~\bibnamefont {Kivelson}},\
  }\href {\doibase 10.1103/PhysRevLett.62.82} {\bibfield  {journal} {\bibinfo
  {journal} {Phys. Rev. Lett.}\ }\textbf {\bibinfo {volume} {62}},\ \bibinfo
  {pages} {82} (\bibinfo {year} {1989})}\BibitemShut {NoStop}%
\bibitem [{\citenamefont {Landau}\ and\ \citenamefont
  {Lifshitz}(1986)}]{Landau1986}%
  \BibitemOpen
  \bibfield  {author} {\bibinfo {author} {\bibfnamefont {L.}~\bibnamefont
  {Landau}}\ and\ \bibinfo {author} {\bibfnamefont {E.}~\bibnamefont
  {Lifshitz}},\ }\href@noop {} {\emph {\bibinfo {title} {Theory of Elasticity,
  3rd}}}\ (\bibinfo  {publisher} {Pergamon Press, Oxford, UK},\ \bibinfo {year}
  {1986})\BibitemShut {NoStop}%
\bibitem [{\citenamefont {Moore}\ and\ \citenamefont {Read}(1991)}]{Read1991}%
  \BibitemOpen
  \bibfield  {author} {\bibinfo {author} {\bibfnamefont {G.}~\bibnamefont
  {Moore}}\ and\ \bibinfo {author} {\bibfnamefont {N.}~\bibnamefont {Read}},\
  }\href {\doibase 10.1016/0550-3213(91)90407-O} {\bibfield  {journal}
  {\bibinfo  {journal} {Nucl. Phys. B}\ }\textbf {\bibinfo {volume} {360}},\
  \bibinfo {pages} {362} (\bibinfo {year} {1991})}\BibitemShut {NoStop}%
\bibitem [{\citenamefont {Haldane}(1983)}]{Haldane1983}%
  \BibitemOpen
  \bibfield  {author} {\bibinfo {author} {\bibfnamefont {F.~D.~M.}\
  \bibnamefont {Haldane}},\ }\href {\doibase 10.1103/PhysRevLett.51.605}
  {\bibfield  {journal} {\bibinfo  {journal} {Phys. Rev. Lett.}\ }\textbf
  {\bibinfo {volume} {51}},\ \bibinfo {pages} {605} (\bibinfo {year}
  {1983})}\BibitemShut {NoStop}%
\bibitem [{\citenamefont {Wen}\ and\ \citenamefont {Zee}(1992)}]{WenZee1992}%
  \BibitemOpen
  \bibfield  {author} {\bibinfo {author} {\bibfnamefont {X.~G.}\ \bibnamefont
  {Wen}}\ and\ \bibinfo {author} {\bibfnamefont {A.}~\bibnamefont {Zee}},\
  }\href {\doibase 10.1103/PhysRevLett.69.953} {\bibfield  {journal} {\bibinfo
  {journal} {Phys. Rev. Lett.}\ }\textbf {\bibinfo {volume} {69}},\ \bibinfo
  {pages} {953} (\bibinfo {year} {1992})}\BibitemShut {NoStop}%
\bibitem [{\citenamefont {Bernevig}\ and\ \citenamefont
  {Regnault}(2009)}]{Bernevig2009}%
  \BibitemOpen
  \bibfield  {author} {\bibinfo {author} {\bibfnamefont {B.~A.}\ \bibnamefont
  {Bernevig}}\ and\ \bibinfo {author} {\bibfnamefont {N.}~\bibnamefont
  {Regnault}},\ }\href {\doibase 10.1103/PhysRevLett.103.206801} {\bibfield
  {journal} {\bibinfo  {journal} {Phys. Rev. Lett.}\ }\textbf {\bibinfo
  {volume} {103}},\ \bibinfo {pages} {206801} (\bibinfo {year}
  {2009})}\BibitemShut {NoStop}%
\bibitem [{\citenamefont {Rezayi}\ and\ \citenamefont
  {Haldane}(1994)}]{Haldane1994}%
  \BibitemOpen
  \bibfield  {author} {\bibinfo {author} {\bibfnamefont {E.}~\bibnamefont
  {Rezayi}}\ and\ \bibinfo {author} {\bibfnamefont {F.}~\bibnamefont
  {Haldane}},\ }\href {\doibase 10.1103/PhysRevB.50.17199} {\bibfield
  {journal} {\bibinfo  {journal} {Phys. Rev. B}\ }\textbf {\bibinfo {volume}
  {50}},\ \bibinfo {pages} {17199} (\bibinfo {year} {1994})}\BibitemShut
  {NoStop}%
\bibitem [{\citenamefont {Luttinger}(1960)}]{Luttinger1960}%
  \BibitemOpen
  \bibfield  {author} {\bibinfo {author} {\bibfnamefont {J.~M.}\ \bibnamefont
  {Luttinger}},\ }\href {\doibase 10.1103/PhysRev.119.1153} {\bibfield
  {journal} {\bibinfo  {journal} {Phys. Rev.}\ }\textbf {\bibinfo {volume}
  {119}},\ \bibinfo {pages} {1153} (\bibinfo {year} {1960})}\BibitemShut
  {NoStop}%
\bibitem [{\citenamefont {Haldane}(1993)}]{Haldane1993}%
  \BibitemOpen
  \bibfield  {author} {\bibinfo {author} {\bibfnamefont {F.~D.~M.}\
  \bibnamefont {Haldane}},\ }\href@noop {} {\emph {\bibinfo {title}
  {Proceedings of the International School of Physics "Enrico Fermi", Course
  CXXI "Perspectives in Many-Particle Physics" : Luttinger's theorem and
  Bosonization of the Fermi surface}}}\ (\bibinfo  {publisher}
  {north-holland},\ \bibinfo {address} {Amsterdam},\ \bibinfo {year} {1993})\
  \Eprint {http://arxiv.org/abs/cond-mat/0505529} {arXiv:cond-mat/0505529}
  \BibitemShut {NoStop}%
\bibitem [{\citenamefont {Varjas}\ \emph {et~al.}(2013)\citenamefont {Varjas},
  \citenamefont {Zaletel},\ and\ \citenamefont {Moore}}]{varjs}%
  \BibitemOpen
  \bibfield  {author} {\bibinfo {author} {\bibfnamefont {D.}~\bibnamefont
  {Varjas}}, \bibinfo {author} {\bibfnamefont {M.~P.}\ \bibnamefont {Zaletel}},
  \ and\ \bibinfo {author} {\bibfnamefont {J.~E.}\ \bibnamefont {Moore}},\
  }\href {\doibase 10.1103/PhysRevB.88.155314} {\bibfield  {journal} {\bibinfo
  {journal} {Phys. Rev. B}\ }\textbf {\bibinfo {volume} {88}},\ \bibinfo
  {pages} {155314} (\bibinfo {year} {2013})}\BibitemShut {NoStop}%
\bibitem [{\citenamefont {Estienne}\ \emph {et~al.}(2013)\citenamefont
  {Estienne}, \citenamefont {Papi\ifmmode~\acute{c}\else \'{c}\fi{}},
  \citenamefont {Regnault},\ and\ \citenamefont {Bernevig}}]{Regnault2013}%
  \BibitemOpen
  \bibfield  {author} {\bibinfo {author} {\bibfnamefont {B.}~\bibnamefont
  {Estienne}}, \bibinfo {author} {\bibfnamefont {Z.}~\bibnamefont
  {Papi\ifmmode~\acute{c}\else \'{c}\fi{}}}, \bibinfo {author} {\bibfnamefont
  {N.}~\bibnamefont {Regnault}}, \ and\ \bibinfo {author} {\bibfnamefont
  {B.~A.}\ \bibnamefont {Bernevig}},\ }\href {\doibase
  10.1103/PhysRevB.87.161112} {\bibfield  {journal} {\bibinfo  {journal} {Phys.
  Rev. B}\ }\textbf {\bibinfo {volume} {87}},\ \bibinfo {pages} {161112}
  (\bibinfo {year} {2013})}\BibitemShut {NoStop}%
\bibitem [{\citenamefont {Haldane}\ and\ \citenamefont {Park}()}]{newpaper}%
  \BibitemOpen
  \bibfield  {author} {\bibinfo {author} {\bibfnamefont {F.~D.~M.}\
  \bibnamefont {Haldane}}\ and\ \bibinfo {author} {\bibfnamefont
  {Y.}~\bibnamefont {Park}},\ }\href@noop {} {\ }\bibinfo {note}
  {Unpublished}\BibitemShut {NoStop}%
\end{thebibliography}
